\documentclass[traditabstract]{aa}

\usepackage{graphicx}
\usepackage{txfonts}
\usepackage{hyperref}

\usepackage{amsmath}
\usepackage{amssymb}
\usepackage{graphicx}
\usepackage{newtxtext,newtxmath}
\usepackage{Times}

\usepackage[T1]{fontenc}
\usepackage{ae,aecompl}

\usepackage{booktabs}
\usepackage{lipsum}
\usepackage{multirow}
\usepackage{natbib}
\usepackage{soul}

\newcommand{\citeprep}[1]{{\color{blue}#1}}
\newcommand{\msun}[1]{$10^{#1}\mathrm{M_\odot}$}

\makeatletter
\newcommand*{\citelink}[1]{\hyper@link{cite}{cite.#1}}

\makeatother
\newcommand{\citeAPA}{\citelink{paulino-afonso2018a}{PA18}}

\bibpunct[ ]{(}{)}{;}{a}{}{,}

\hypersetup{
     colorlinks   = True,
     citecolor    = blue,
     linkcolor = blue,
     urlcolor = blue
}

\begin{document} 

%%%%%%%%%%%%%%%%%%%%%
%%%%%%%% TITLE PAGE %%%%%%
%%%%%%%%%%%%%%%%%%%%%

\title{VIS$\boldsymbol{^3}$COS: II. Nature and nurture in galaxy structure and morphology}

\author{
Ana Paulino-Afonso \inst{1,2,3}\fnmsep\thanks{E-mail: aafonso@oal.ul.pt} \and
David Sobral \inst{3}
\and
Behnam Darvish \inst{4} 
\and 
Bruno Ribeiro \inst{5}
\and
Arjen van der Wel\inst{6,7}
\and
John Stott \inst{3}
\and
Fernando Buitrago\inst{1,2}
\and
Philip Best\inst{8}
\and
Andra Stroe \inst{9}
\and 
Jessica E. M. Craig \inst{3}
}

\institute{Instituto de Astrof\'isica e Ci\^encias do Espa\c{c}o, Universidade de Lisboa, OAL, Tapada da Ajuda, PT1349-018 Lisboa, Portugal
\and
Departamento de F\'isica, Faculdade de Ci\^encias, Universidade de Lisboa, Edif\'icio C8, Campo Grande, PT1749-016 Lisboa, Portugal
\and
Department of Physics, Lancaster University, Lancaster, LA1 4YB, UK
\and
Cahill Center for Astrophysics, California Institute of Technology, 1216 East California Boulevard, Passadena, CA 91125, USA
\and
Leiden Observatory, Leiden University, P.O. Box 9513, NL-2300 RA Leiden, The Netherlands
\and
Max-Planck-Institut f\"ur Astronomie, K\"onigstuhl 17, D-69117, Heidelberg, Germany
\and
Sterrenkundig Observatorium, Universiteit Gent, Krijgslaan 281 S9, B-9000 Gent, Belgium
\and
Institute for Astronomy, University of Edinburgh, Royal Observatory, Blackford Hill, Edinburgh EH9 3HJ, UK
\and
Harvard-Smithsonian Center for Astrophysics, 60 Garden Street, Cambridge, MA 02138, USA
}

\date{ }
\date{Received 28 January 2019; accepted 19 July 2019}

\abstract{
We study the impact of local density and stellar mass on the structure and morphology of approximately 500 quiescent and star-forming galaxies from the VIMOS Spectroscopic Survey of a Superstructure in COSMOS (VIS$^{3}$COS). We perform bulge-to-disc decomposition of the surface brightness profiles and find $\sim41\pm3$\% of $>10^{10}\mathrm{M_\odot}$ galaxies to be best fitted with two components. We complement our analysis with non-parametric morphological measurements and qualitative visual classifications. We find that both galaxy structure and morphology depend on stellar mass and environment for our sample as a whole. We only find an impact of the environment on galaxy size for galaxies more massive than \msun{11}. We find higher S\'ersic indices ($n$) and bulge-to-total ratios ($B/T$) in high-density regions when compared to low-density counterparts at similar stellar masses. We also find that galaxies with higher stellar mass have steeper light profiles (high $n$, $B/T$) compared to galaxies with lower stellar mass. Using visual classifications, we find a morphology--density relation at $z\sim0.84$ for galaxies more massive than $10^{10}\mathrm{M_\odot}$, with elliptical galaxies being dominant at high-density regions and disc galaxies more common in low-density regions. However, when splitting the sample into colour--colour-selected star-forming and quiescent sub-populations, there are no statistically significant differences between low- and high-density regions. We find that quiescent galaxies are smaller, have higher S\'ersic indices (for single profiles, around $n\sim4$), and higher bulge-to-total light ratios (for decomposed profiles, around $B/T\sim0.5$) when compared to star-forming counterparts ($n\sim1$ and $B/T\sim0.3$, for single and double profiles, respectively). We confirm these trends with non-parametric quantities, finding quiescent galaxies to be smoother (lower asymmetry, lower $M_{20}$) and to have most of their light over smaller areas (higher concentration and Gini coefficient) than star-forming galaxies. Overall, we find a stronger dependence of structure and morphology on stellar mass than on local density and these relations are strongly correlated with the quenching fraction. The change in average structure or morphology corresponds to a change in the relative fractions of blue disc-like galaxies and red elliptical galaxies with stellar mass and environment. We hypothesise that the processes responsible for the quenching of star formation must also affect the galaxy morphology on similar timescales.
\vspace*{0.0cm}
}

\keywords{galaxies: evolution -- galaxies: high-redshift -- galaxies: structure -- large-scale structure of Universe}

\authorrunning{A. Paulino-Afonso et al.}
\titlerunning{Nature and nurture in galaxy structure and morphology}

\maketitle

%%%%%%%%%%%%%%%%%%%%%
%%%%%%% SECTION 1 %%%%%%%
%%%%%%%%%%%%%%%%%%%%%

\section{Introduction}\label{section:introduction}

In a $\Lambda$CDM universe, galaxies form in dark matter halos when baryonic matter cools and collapses \citep[e.g.][]{white1978}. This provides a hierarchical scenario where massive objects are formed through mergers of smaller entities. However, the exact details of galaxy formation and evolution still elude our current understanding. The hierarchical nature of structure formation naturally produces different pathways of galaxy evolution based on the local density, as denser regions have a higher probability of interactions that influence galaxy properties. 

By studying samples of galaxies across different regions, \citet{dressler1980} found a clear dichotomy in galaxy morphology when looking at low- (hereafter referred as field) and high-density (cluster) environments in the local Universe \citep[see also e.g.][]{guzzo1997,goto2003,bamford2009,skibba2009,fasano2015}. Galaxies in field environments are on average bluer, more star-forming, and disc-like while galaxies in cluster environments are older, redder, less star-forming, and elliptical \citep[e.g.][]{dressler1984,gomez2003,kauffmann2004,boselli2006,blanton2009,deeley2017}. 

Changes in galaxy morphology with environment are not only found in the local Universe but also at intermediate \citep[$z \lesssim 1$, e.g.][]{dressler1997,treu2003,postman2005,capak2007,vanderwel2007,tasca2009,kovac2010,nantais2013b,allen2016,krywult2017,kuchner2017} and high redshifts \citep[$z\sim 1-2$, e.g.][]{grutzbauch2011,bassett2013,strazzullo2013,allen2015}. There are some hints of the environmental impact on galaxy size at $z\sim 1-2$ \citep[e.g.][]{papovich2012,delaye2014,mei2015} but that is not seen in the local Universe \citep[e.g.][]{huertas-company2013, kelkar2015} or in protocluster environments \citep[see e.g.][]{peter2007}. By measuring  sizes of field and cluster galaxies, several studies find quiescent galaxies to show little difference in their sizes at fixed stellar mass at $0<z<2$ \citep[e.g.][]{huertas-company2013b,huertas-company2013,cebrian2014,newman2014,kelkar2015,allen2015,allen2016,saracco2017} while others find evidence for larger quiescent galaxies in cluster environments \citep[e.g.][]{papovich2012,bassett2013,lani2013, strazzullo2013, delaye2014,yoon2017}. For star-forming galaxies, there is also not a clear trend, with some studies finding little difference among cluster and field galaxies \citep[e.g.][]{lani2013,kelkar2015} and others finding larger star-forming galaxies in cluster environments \citep[e.g.][\!\!, locally and at $z\sim2$, respectively]{cebrian2014,tran2017}. Studies by \citet{allen2015,allen2016} show that star-forming galaxies are larger in cluster environments at $z\sim$1 and smaller at $z\sim$2 than their field counterparts. This differential evolution of galaxies of different sizes hints at different paths for galaxy growth in different environments.

Differences among star-forming and quiescent galaxies can evolve through the morphological transformation of blue star-forming disc-dominated galaxies to redder quiescent and bulge-dominated (or pure elliptical) galaxies \citep[e.g. through minor and major mergers,][]{delucia2011,shankar2014}. In terms of galaxy light profiles, studies find that galaxies residing in the cluster environments might be more bulge-dominated \citep[e.g.][]{goto2003,poggianti2008,skibba2012,bluck2014}. By quantifying the light distribution in galaxies with \citet{sersic1968} profiles, \citet{allen2016} find that in and around a $z\sim0.92$ cluster, star-forming galaxies are more likely to have higher S\'ersic indices than their field counterparts but report no difference among quiescent galaxies. At $z\sim1.6$, \citet{bassett2013} find no differences between field and cluster star-forming galaxies but report shallower profiles (lower S\'ersic index) for quiescent galaxies in a cluster environment. When comparing star-forming and quiescent galaxies, the latter have higher S\'ersic indices due to a prevalence of ellipticals and/or a dominant bulge in the redder population \citep[e.g.][]{bassett2013,morishita2014,cerulo2017}. Galaxies with a high S\'ersic index are also the types of galaxies that are more common in higher-density regions out to $z\sim1$ \citep[e.g.][]{dressler1997,treu2003,postman2005,capak2007,vanderwel2007,tasca2009,nantais2013b}.

When performing more detailed bulge-to-disc decomposition of the light profiles, studies find a rise in the bulge-dominated fraction from $z\sim3$ \citep[e.g.][]{bruce2014b,tasca2014,margalef-bentabol2016}. In the local Universe, there are hints that the build-up of galactic bulges is happening in higher-density environments \citep[e.g.][]{lackner2013}. At intermediate redshifts ($z\sim$0.4-0.8), \citet[][]{grossi2018} find that a sample of H$\alpha$-selected galaxies tend to have more prominent bulges in higher-density environments. However, we lack observations of the environmental dependence of the bulge prevalence at these redshifts for a continuum-selected sample.

The morphology--colour--density relation suggests that there is at least one physical mechanism that changes galaxy morphology and also acts on the suppression of star formation activity. Several processes have been proposed, including gas removal from the disc \citep[e.g.][]{larson1980}, ram pressure stripping from the intra-cluster medium \citep[e.g.][]{gunn1972,abramson2016}, galaxy harassment through tidal forces \citep[e.g.][]{moore1996}, and eventual galaxy mergers \citep[e.g.][]{burke2013}. At the same time, there is a typical stellar mass in which quenching is effective due to overdense environments \citep[e.g.][]{peng2010b,peng2012}.

Here, we study a sample of spectroscopically confirmed sources in and around a superstructure at $z\sim0.84$ in the COSMOS field \citep{scoville2007} for which we have available high-resolution spectra covering [O{\sc ii}], the $4000$\AA\ break, and H$\delta$  \citep[][\!\!, hereafter \citeprep{PA18}]{paulino-afonso2018}. We aim to investigate the relationship between galaxy morphology and stellar mass and environment, and link that to the star formation to shed some light on the processes that are most likely to be responsible for morphological transformations.

This paper is organised as follows: in Section \ref{section:data} we briefly explain the VIMOS Spectroscopic Survey of a Superstructure in the COSMOS field (VIS$^3$COS; \citeAPA), on which our study is based. Section \ref{section:methods} details the morphological measurements on the sources used in this article. In Sections \ref{section:results_sm} and \ref{section:results_sfr} we highlight some of the key results of our study in terms of galaxy stellar mass, environment, and star formation. In Section \ref{section:discussion} we discuss our findings within the context of current galaxy formation and evolution literature. In Section \ref{section:conclusions} we summarise our results. We use AB magnitudes \citep{oke1983}, a Chabrier \citep{Chabrier2003} initial mass function (IMF), and assume a $\Lambda$CDM cosmology with H$_{0}$=70 km s$^{-1}$Mpc$^{-1}$, $\Omega_{M}$=0.3, and $\Omega_{\Lambda}$=0.7. The physical scale at the redshift of the superstructure ($z\sim0.84$) is 7.63 kpc/\arcsec.

%%%%%%%%%%%%%%%%%%%%%
%%%%%%%% SECTION 2 %%%%%%
%%%%%%%%%%%%%%%%%%%%%

\section{Sample and data}\label{section:data}

        \subsection{The VIS$^3$COS survey}

\begin{figure*}
\centering
\includegraphics[width=\linewidth]{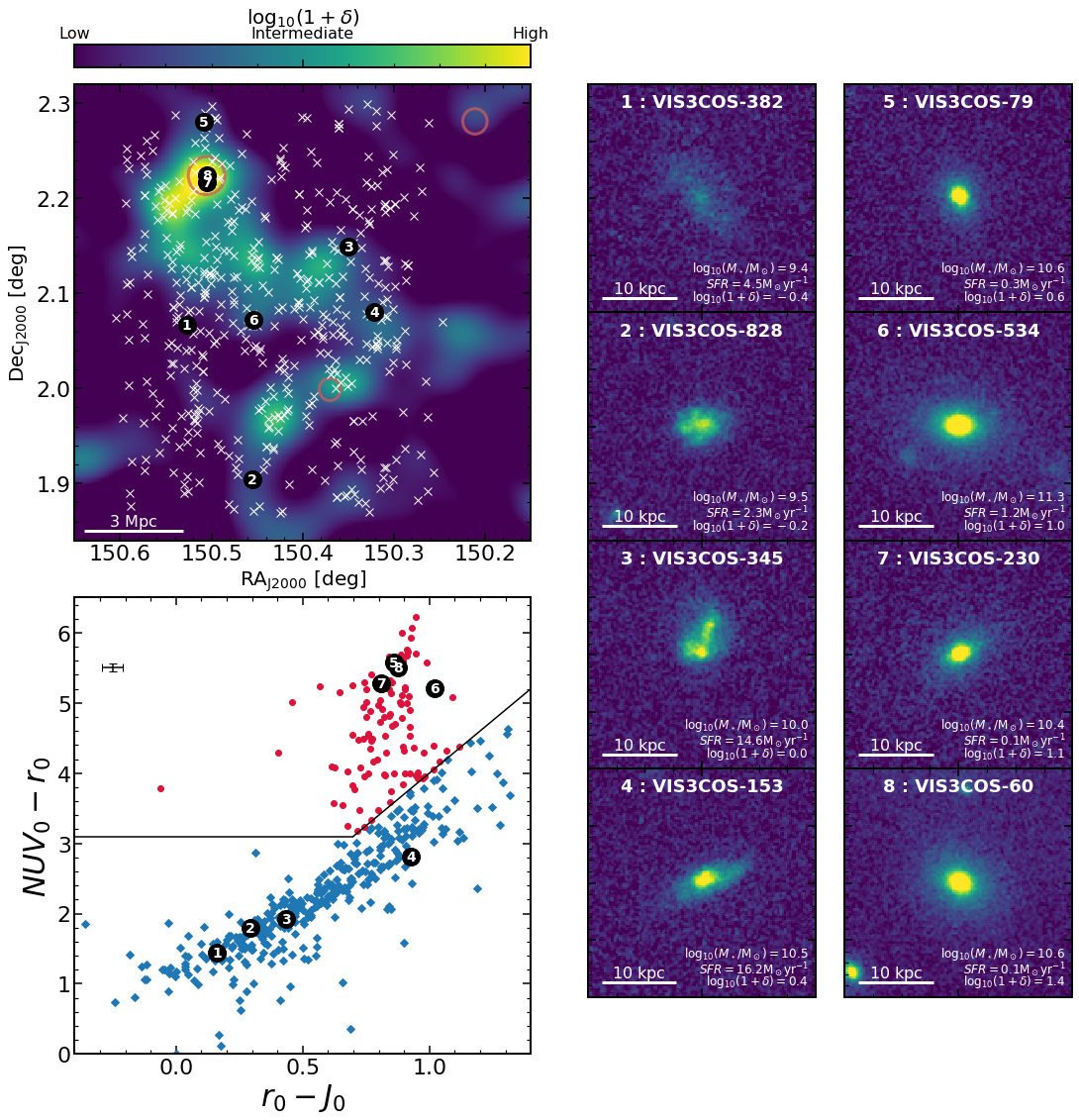}
\caption{Top left: Overview of the VIS$^3$COS survey showing the galaxy overdensity and targeted galaxies at $0.8<z<0.9$ with spectroscopic redshifts (white crosses) along with the location of known X-ray clusters \citep[empty red circles,][]{finoguenov2007}. Bottom left: NUV-r-J diagram (derived using \citealt{laigle2016} photometry) for galaxies in our survey, with the separation between quiescent (red circles) and star-forming (blue diamonds) as defined by \citet{ilbert2013} shown as a solid line. We show the average error on each colour as a black cross. Right panels: Examples of HST/ACS F814W 4\arcsec$\times$4\arcsec\ rest-frame $B$-band images \citep{koekemoer2007} of eight of our sources with individual information on stellar mass, SFR, and local overdensity in each panel. We highlight the position of these eight galaxies with large numbered black circles in the left panels.} 
\label{fig:surveySummary}
\end{figure*}

The VIS$\boldsymbol{^3}$COS survey is based on an observing programme with the VIMOS \footnotemark{}\footnotetext{Programmes 086.A-0895, 088.A-0550, and 090.A-0401} instrument mounted at the VLT to obtain high-resolution spectroscopy down to the continuum level for galaxies in and around a large structure at $z\sim0.84$ in the COSMOS field. The observations span an area of 21\arcmin$\times$31\arcmin\ (9.6$\times$14.1 Mpc) with an overdensity of H$\alpha$ emitters \citep{sobral2011,darvish2014} and three confirmed X-ray clusters \citep{finoguenov2007}. This is the third publication from this survey and the full description of the data and derived physical quantities is presented in \citeprep{PA18}. We summarise the relevant information below.

We targeted galaxies from the \citet{ilbert2009} catalogue which had $0.6<z_\mathrm{phot,l}<1.0$ (with $z_\mathrm{phot,l}$ being either the upper or lower 99\% confidence interval limit for each source) and were brighter than $i_\mathrm{AB}<23$. We used the VIMOS high-resolution red grism (with the GG475 filter, $R\sim2500$) with six overlapping VIMOS pointings to mitigate selection effects on higher-density regions. Our choice of grism covers the 3400-4600\,\AA\ rest-frame region at the redshift of the superstructure, which has interesting spectral features such as [O{\sc ii}]\,$\lambda$3726,$\lambda$3729 (partially resolved doublet), the 4000\,\AA\ break, and H$\delta$.

The spectroscopic redshifts were computed from the extracted 1D spectra using \textsc{SpecPro} \citep{masters2011}. The redshift determination is based on a set of prominent spectral features: [O{\sc ii}], H+K absorption, G-band, some Fe lines, and H$\delta$. We obtained successful spectroscopic redshifts for 696 galaxies, of which 490 are within our primary redshift selection ($0.8<z<0.9$, \citeprep{PA18}).

With the knowledge of the spectroscopic redshift, we can improve on existing physical quantity measurements. We obtained stellar masses and star formation rates (SFRs) from running \textsc{magphys} \citep{cunha2008} with spectral models constructed from the stellar libraries by \citet{bruzual2003} on the set of photometric bands from near-UV to near-IR from the COSMOS2015 catalogue \citep{laigle2016}. The dust is modelled based on the \citet{charlot2000} prescription.

We use a measurement of local overdensity based on the cosmic density field value at the 3D position of each target. We use the density estimation of \citet{darvish2015a,darvish2017} which is constructed from a $K_s$ magnitude-limited sample based on the \citet{ilbert2013} photometric redshift catalogue. The density field was computed for the $\sim1.8\,\mathrm{deg^{2}}$ area in COSMOS over a large redshift interval ($0.05<z_\mathrm{phot}<3.2$) with an adaptive smooth kernel with a characteristic size of 0.5 Mpc \citep{darvish2015a,darvish2017}. In this manuscript, we define overdensity as $1+\delta=\Sigma/\Sigma_\mathrm{median}$, with $\Sigma_\mathrm{median}$ being the median of the density field at the redshift of the galaxy. For a detailed description of the density estimation method, we refer the reader to \citet{darvish2015a,darvish2017}.

The final sample we study in this manuscript is selected to be at $0.8<z<0.9$ (matching our target selection) and has a total of 490 galaxies spanning a large diversity of stellar masses (with 295 above our selection limit $\sim10^{10}\mathrm{M_\odot}$, \citeprep{PA18}) and environments across $\sim$10 Mpc. We show an overview of the main properties of the sample and survey in Fig. \ref{fig:surveySummary}. We also note that we are probing both star-forming (371 galaxies) and quiescent (119 galaxies) populations within this region \citep[defined from the NUV-r-J diagram; see e.g.][{and Fig. \ref{fig:surveySummary}}]{ilbert2013}.

        \subsection{Imaging data}

Since this structure is part of the COSMOS field, we base our morphological measurements on data from the HST/ACS F814W COSMOS survey \citep[][]{koekemoer2007,scoville2007}. These images have a typical PSF FWHM of $\sim 0.09$\arcsec, a pixel scale of $0.03\arcsec/\mathrm{pixel}$, and a limiting point-source depth AB(F814W) = 27.2 (5 $\sigma$). At the redshift of the superstructure, these images probe the rest-frame $B$-band galaxy morphology with sub-kiloparsec resolution.

We use 10\arcsec$\times$10\arcsec cut-outs (corresponding to square images with a $\sim76$ kpc side at the redshift of the superstructure) centred on the target position. To account for the PSF, we use the HST/ACS PSF profiles that were created with \textsc{TinyTim} \citep{krist1995} models and described by {\citet[][\!; see also \citealt{paulino-afonso2017}]{rhodes2006,rhodes2007}}.

%%%%%%%%%%%%%%%%%%%%%
%%%%%%%% SECTION 3 %%%%%%
%%%%%%%%%%%%%%%%%%%%%

\section{Morphological characterisation of the sample}\label{section:methods}

Quantitative morphological analysis has complemented visual classification of images in the past few decades. There are two main groups of morphological characterisation: parametric modelling of the surface brightness profiles \citep[e.g.][]{vaucouleurs1959,sersic1968,simard1998,trujillo2001,souza2004,peng2002,peng2010,simard2011} and non-parametric quantitative morphology \citep[e.g.][]{abraham1994,abraham2003,bershady2000, conselice2000,conselice2003,papovich2003,lotz2004,blakeslee2006,law2007,freeman2013,pawlik2016}. Each method has its own strengths and weaknesses and a choice between the two is usually related to a particular scientific question. Parametric models are more effective in obtaining a description of the light profile to get galaxy size estimates \citep[e.g.][]{blanton2003,trujillo2007,buitrago2008,wuyts2011,vanderwel2014} and to perform bulge-to-disc decomposition \citep[e.g.][]{souza2004,tasca2009,simard2011,meert2013,bruce2014b,bruce2014a,lang2014,margalef-bentabol2016,gao2017,dimauro2018}. Non-parametric methods are often used to identify irregularities in galaxies as signatures of past or ongoing mergers \citep[e.g.][]{lotz2008,conselice2009,bluck2012,freeman2013,pawlik2016}. Since we are interested in the process of morphological transformation from low to dense environments, we use a combination of both methods along with visual classification in order to have a complete perspective on the impact of environment on galaxy morphology.

        \subsection{Parametric modelling of galaxies}\label{ssection:parametric_models}

To obtain an estimate of the structural parameters of galaxies we fitted \citet{sersic1968} profiles to all objects in our catalogue using \textsc{GALFIT} \citep{peng2002,peng2010}. We also use \textsc{SExtractor} \citep{bertin1996} to provide initial guesses for each galaxy model and to produce binary images to mask all nearby objects that might affect the fit. This method closely follows the procedures defined in \citet{paulino-afonso2017} and \citet{paulino-afonso2018b}. We fitted all galaxies with two models: a single S\'ersic profile and a combination of an exponential disc with a central S\'ersic profile to account for the existence of a bulge+disc system. We chose to do so since we are dealing with a population of galaxies where substructures can be resolved \citep[see e.g.][]{tasca2009}. We use the Bayesian information criterion \citep[BIC, e.g.][]{kelvin2012proc,bruce2014a} to select which model best fits each galaxy (see Appendix \ref{app:bic} for more details).

        \subsection{Non-parametric quantitative morphology}\label{ssec:nonpar}

We implement two sets of non-parametric indices that allow us to get additional structural indicators without the need to assume any model: the CAS system \citep[][\!\!, see also \citealt{abraham1994} and \citealt{bershady2000}]{conselice2000,conselice2003} and the G-$M_{20}$ system \citep[][\!, see also \citealt{abraham2003}]{lotz2004}. The two latter indices are computed over the segmentation map of the galaxy, which is computed as the group of a minimum of ten connected pixels above 3$\sigma$ that are closest to the object coordinates. These indices are commonly used to detect disturbed galaxy light profiles associated with ongoing galaxy mergers \citep[e.g.][]{conselice2003,lotz2004,lotz2008,conselice2009}. For more details on each index see Appendix \ref{app:np}.

        \subsection{Visual classification}\label{ssection:visual_class}

The classification of galaxies into different categories has been done extensively over a century \cite[e.g.][]{hubble1926,hubble1930, vaucouleurs1959,vandenbergh1976,nair2010,baillard2011,buitrago2013,buta2015,kartaltepe2015}. This is a time-consuming task if one wishes to carry it out on large samples; it is also not reproducible and is subject to individual bias. More recently, the citizen science project Galaxy Zoo \citep{lintott2008} combined results from more than 200,000 classifiers to produce a reliable catalogue of visual classifications \citep{lintott2011,willett2013,willett2017}. In this manuscript, we use the data release of Galaxy Zoo containing the classifications for Hubble Space Telescope images, fully described by \citet{willett2017}. Out of 490 galaxies within our sample at $0.8<z<0.9$ we find a match for 447 objects. To map the classifications from Galaxy Zoo to the three classical morphologies \citep[elliptical, disc, or irregular, see e.g.][]{paulino-afonso2018b}, we use the first and second tier questions \citep[][\!\!, Figure 4]{willett2017}. We use the recommended fractions (with corrections for classification bias) and establish the following criteria (for more details see Appendix \ref{app:visualclass}):
\begin{itemize}
\item Elliptical -- $f_\mathrm{smooth}>0.50$ and $f_\mathrm{odd}<0.5$ and $f_\mathrm{cigar-shaped}<0.5$ and $f_\mathrm{features}<0.23$ (to impose mutually exclusive classification);
\item Disc -- $f_\mathrm{features}>0.23$\footnotemark{}\footnotetext{As suggested by \citet[][\!\!, see Table 11]{willett2017} when considering fractions on the second tier of questions.},  and $f_\mathrm{clumpy}<0.5$ and $f_\mathrm{odd}<0.5$ or $f_\mathrm{smooth}>0.50$ and $f_\mathrm{odd}<0.5$ and $f_\mathrm{cigar-shaped}>0.5$;
\item Irregular -- $f_\mathrm{odd}>0.5$ or  $f_\mathrm{features}>0.23$ and $f_\mathrm{clumpy}>0.5$.
\end{itemize}

%%%%%%%%%%%%%%%%%%%%%
%%%%%%%% SECTION 4 %%%%%%
%%%%%%%%%%%%%%%%%%%%%

\begin{figure*}
\centering
\includegraphics[width=0.85\linewidth]{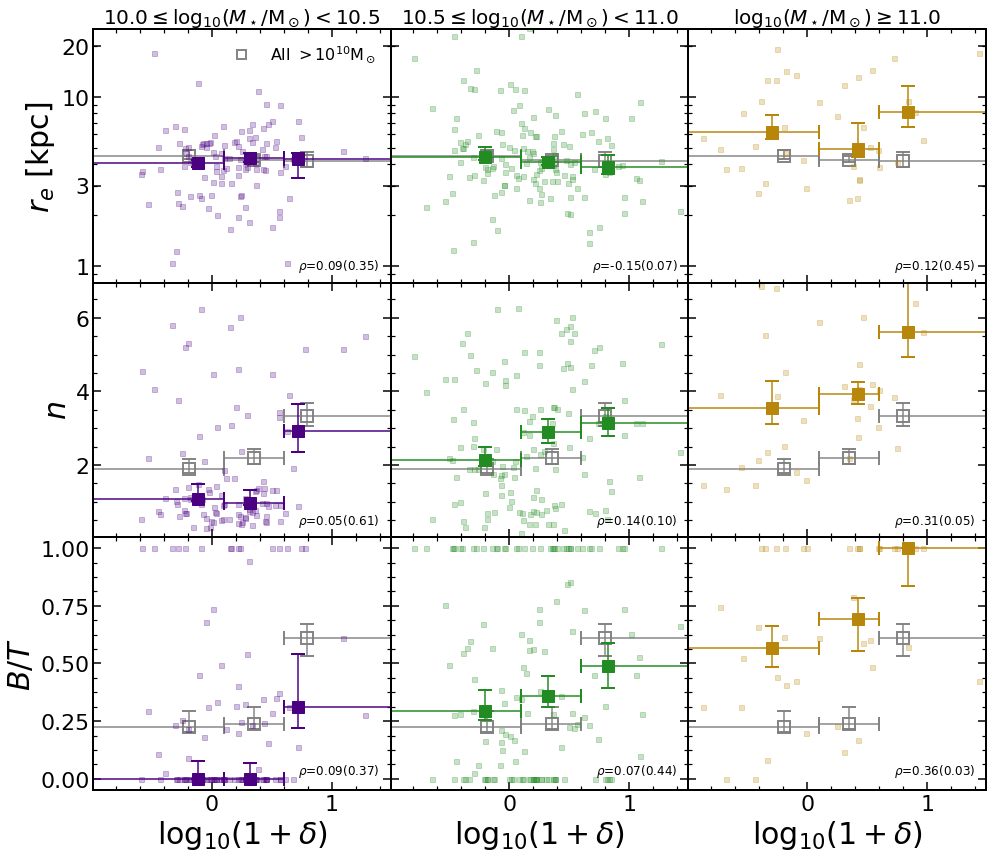}
\caption{Dependence of $z\sim0.84$ galaxy sizes (top), S\'ersic indices (middle), and the bulge-to-total ratio (bottom) on the environment for three different stellar mass bins (from left to right). We add to all panels the relation for the global sample at $M_\star>$\msun{10} as empty grey squares. In each panel, we show the Spearman correlation coefficient, $\rho$, and the corresponding probability of an uncorrelated dataset having the same distribution in parentheses. We find sizes are roughly constant across different environments but increase with stellar mass. In terms of their light profiles, we see a trend where both stellar mass and environment have an impact with more massive galaxies and denser environments showing larger values for $n$ and $B/T$.}
\label{fig:parametric_mass_env}
\end{figure*} 

\begin{figure*}
\centering
\includegraphics[width=0.85\linewidth]{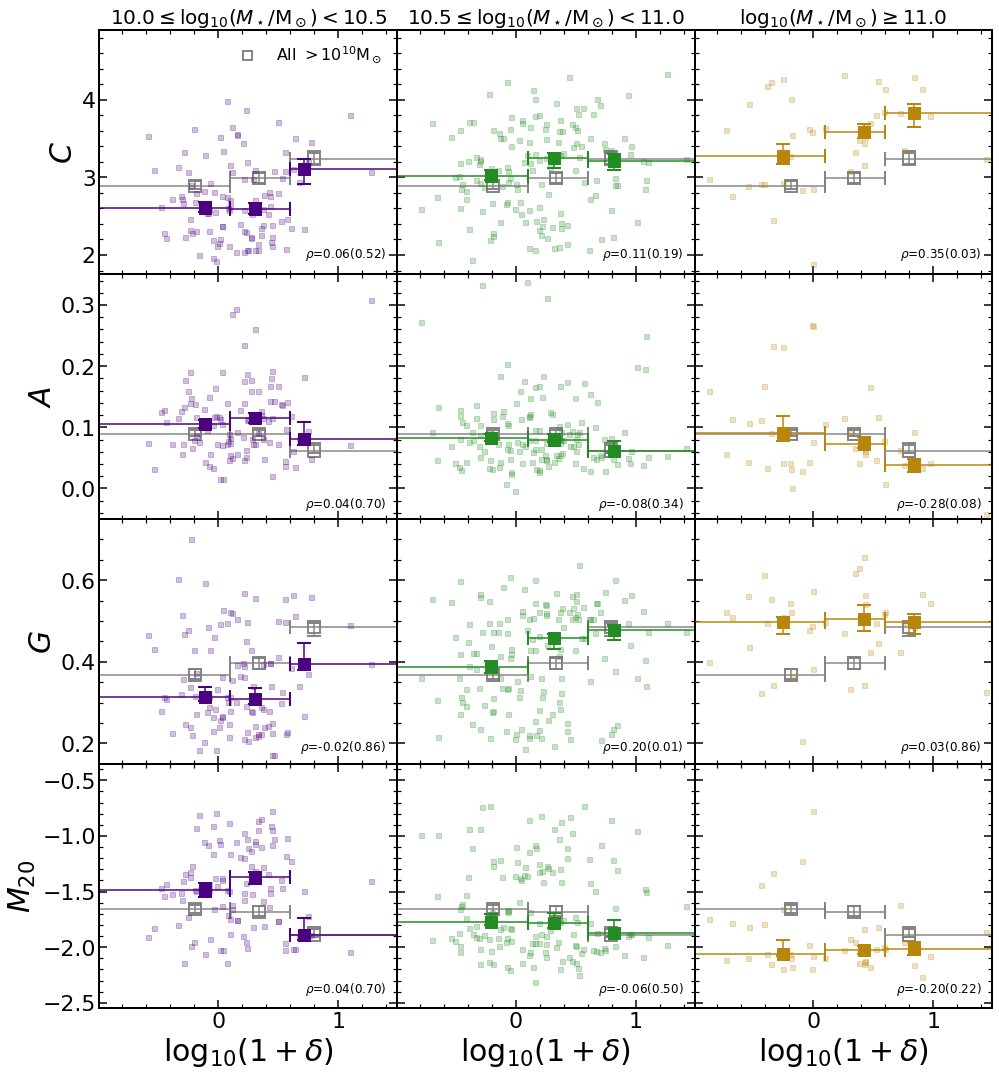}
\caption{Dependence of non-parametric tracers (from top to bottom: light concentration, asymmetry, Gini, and moment of light) on the environment for three different stellar mass bins (from left to right). We add to all panels the relation for the global sample at $M_\star>$\msun{10} as empty grey squares. In each panel, we show the Spearman correlation coefficient, $\rho$, and the corresponding probability of an uncorrelated dataset having the same distribution in parentheses. Overall we find that both stellar mass and environment have some impact on non-parametric morphology, with stellar mass having the strongest impact on the median of the population (as measured by the average gradient)}.
\label{fig:nonparametric_mass_env}
\end{figure*} 

\section{ Dependence of galaxy structure and morphology on stellar mass and environment}\label{section:results_sm}

We group galaxies into three different samples based on the local density in order to trace objects which should be representative of field ($\log_{10}(1+\delta)\leq0.1$), intermediate-density and filaments ($0.1<\log_{10}(1+\delta)\leq0.6$), and cluster galaxies ($\log_{10}(1+\delta)>0.6$) based on the relation of the cosmic web environment with overdensity (see \citeprep{PA18}). Unless stated otherwise, the horizontal error bars delimit the bins and the vertical error bars show the value of the 16th (lower uncertainty) and 84th (upper uncertainty) percentiles of the distribution for each bin normalized by the bin size as $[P_{16\%},P_{84\%}]/\sqrt{N_\mathrm{gal}}$. We also compute the \citet{spearman1904} correlation coefficient, $\rho$, and the probability of the relation being random for all of the relations explored in this manuscript and show them in the individual panels of each figure.

For 470 (96\%) of the 490 galaxies at $0.8<z<0.9$ we successfully fitted their light profiles with either a one- or two-component model. The remaining 20 galaxies failed to converge. Following Section \ref{ssection:parametric_models} we find a total of 173 galaxies for which their best fit is a two-component model. Considering only galaxies with stellar masses greater than $10^{10}\mathrm{M_\odot}$, we find a fraction of $\sim41\pm3$\% of two-component systems. This is in agreement with the reported two-component model fraction of $35\pm6\%$ at $z\sim1$ by \citet{margalef-bentabol2016} for a sample of $\log_{10}\left(M_\star/\mathrm{M_\odot}\right)>10$ galaxies.

        \subsection{Parametric quantities}\label{ssection:par_results}

To compare the morphology of galaxies across different stellar masses and environments in a consistent way, we use as a size estimate the effective radius of the single S\'ersic model for each galaxy. We also do the same when showing S\'ersic indices. For the bulge-to-total ratio ($B/T$), we use the value from the two-component model for galaxies, which has a statistically better fit (see Appendix \ref{app:bic}). For galaxies that are best fitted with a single S\'ersic profile we assign a value of $B/T$=0 if $n<2.5$ and $B/T$=1 if $n\geq2.5$ \citep[see e.g.][regarding the $n$ threshold]{shen2003,barden2005,cebrian2014,lange2015,kuchner2017}. We note that using a different threshold for the separation \citep[e.g. $n=2$,][]{ravindranath2004} does not qualitatively change our results. An alternative would be to introduce an estimate of B/T (between 0 and 1) based on the value of the best fit S\'ersic index of each galaxy. However, doing so would introduce the underlying correlation of stellar mass and local density with the S\'ersic index on all relations for $B/T$, making it more difficult to interpret the results independently.

We show in Fig. \ref{fig:parametric_mass_env} the dependence of galaxy sizes, S\'ersic indices, and $B/T$ on the environment for galaxies more massive than $10^{10}\mathrm{M_\odot}$. We find that for a given stellar mass range there is no significant change in galaxy size (for the low- and intermediate-mass bins) and the correlation with local density is weak ($\rho<0.15$). For the highest-stellar mass bin we find that in the higher-density regions there is a lack of small galaxies ($\lesssim4$kpc) which drives the median value towards $\sim40$\% larger sizes, but the correlation with density is weak ($\rho=0.12$). The larger sizes of galaxies in the densest regions cannot be explained solely by changes in the mean stellar mass for each density bin. The most likely scenario is that growth through galaxy mergers drives this difference \citep[see e.g.][for local early-type galaxies, which are the dominant population in the high-stellar mass bin in our study]{papovich2012,cappellari2013,yoon2017}. On the other hand, we find that more massive galaxies are larger, as expected from the underlying stellar-mass--size relation \citep[see e.g.][]{franx2008, vanderwel2014, morishita2014, paulino-afonso2017, mowla2019}. 

The median S\'ersic index increases with stellar mass with more massive galaxies having steeper light profiles (higher values of $n$). We also find that for a given stellar mass bin, there is an increase in $n$ for denser environments, more specifically, when comparing the densest with the lowest regions probed with VIS\textsuperscript{3}COS. The lack of disc-like galaxies ($n\lesssim2.5$) at all stellar masses in high-density regions is especially noteworthy. We find that the correlation with local density is stronger for the higher-stellar mass bin in our sample ($\rho=0.31$) and that for the lower-stellar mass bin the correlation is not significant  ($\rho=0.05$).

The trends of $B/T$ with stellar mass and environment are seen in Fig. \ref{fig:parametric_mass_env} with strong differences found among galaxies with different stellar masses at a fixed local overdensity. We also find a significant trend with galaxies in denser environments having higher $B/T$ values for fixed stellar mass. A similar trend for $B/T$ as for the S\'ersic index is seen, which is expected since the presence of a more prominent bulge should also produce a steeper light profile in the galaxy centre. Despite the differences observed in the median values, we do find the correlations with local density to be weak for the low- and intermediate-stellar mass bins ($\rho<0.1$). The stronger correlation is observed when considering  galaxies with higher stellar mass ($\rho=0.36$).

Our results highlight that galaxy morphology changes with the environment (at fixed stellar mass) and changes with stellar mass (at a fixed environment) at $z\sim0.84$. We compute the average gradient of the median value for stellar mass and local density. We find that a variation in stellar mass implies a stronger change on the median morphological parameter when compared to a variation in local overdensity. This can also be seen in Fig. \ref{fig:parametric_mass_all}, where we find stronger correlations with stellar mass for all shown quantities.

        \subsection{Non-parametric quantities} 

We summarise the results on non-parametric morphological tracers as a function of stellar mass and environment in Fig. \ref{fig:nonparametric_mass_env}. We find a clear dependence of the median light concentration on stellar mass, with more massive galaxies being more concentrated. We also find a trend with the environment, in which galaxies in denser environments have higher values of $C$. The correlation of $C$ with local density is the strongest for the higher-stellar mass bin ($\rho=0.35$), being close to non-existent in the lower-stellar mass bin ($\rho=0.06$).

When considering the asymmetry of light profiles (see second row of Fig. \ref{fig:nonparametric_mass_env}), we find little dependence of the median asymmetry on both stellar mass and environment. For low- to intermediate-stellar mass bins the correlation is weak to non-existent ($\rho<0.1$). For the higher-stellar mass bin there is a slightly stronger correlation ($\rho=0.28$) than at lower stellar masses with galaxies in high-density regions being less asymmetric than those in lower-density environments.

The Gini coefficient displays a more interesting set of trends on the median of the population (see third row of Fig. \ref{fig:nonparametric_mass_env}). We find a clear trend with stellar mass, with more massive galaxies having higher Gini values (consistent with higher concentration). Concerning local density, we find that the trend depends on the stellar mass of the population. The lower-stellar mass galaxies are similar in low- and intermediate-density environments, but then the Gini coefficient increases towards denser environments. For galaxies of intermediate stellar mass, we find a slight trend of galaxies having larger Gini values from low- to high-density environments. For the most massive galaxies in the sample, we find no environmental dependence. We note, however, that the correlation with local density is absent for the lower and higher-stellar mass bins ($\rho<0.05$), and only significant at intermediate stellar masses ($\rho=0.2$ but only a $\sim$1\% probability of being an uncorrelated distribution).

In terms of the median moment of light, we find a clear trend with stellar mass, with systems of  higher stellar mass having lower values of $M_{20}$ (less disturbed profiles; see fourth row of Fig. \ref{fig:nonparametric_mass_env}). In terms of environmental dependence, we find no significant dependence of the median for the intermediate- and high-stellar mass bins. For  galaxies of lower stellar mass, there is a drop in the value of $M_{20}$ in the densest regions compared to a roughly constant value at lower densities. We note, however, that all of the correlations with local density are weak ($\rho=0.2$ and $\sim$22\% probability of being an uncorrelated distribution at higher stellar masses) to non-existent ($\rho<0.1$ at lower stellar masses).

%%%%%%%%%%%%%%%%%%%%%
%%%%%%%% SECTION 5 %%%%%%
%%%%%%%%%%%%%%%%%%%%%

\section{Relation of galaxy structure and morphology to star formation}\label{section:results_sfr}

In this section, we explore the influence of star formation activity on galaxy structure and morphology by splitting our sample into star-forming and quiescent populations according to the NUV-r-J colour--colour diagram \citep[e.g.][\!, see also Fig. \ref{fig:surveySummary}]{ilbert2013}. Since these two populations have been found to have different typical morphologies and structural parameters \citep[see e.g.][]{vanderwel2008, vanderwel2014, morishita2014}, we want to quantify possible differences with stellar mass and environment produced by having different mixes of the star-forming and quiescent populations.

        \subsection{Galaxy sizes}\label{ssection:sizes}

\begin{figure*}
\centering
\includegraphics[width=0.85\linewidth]{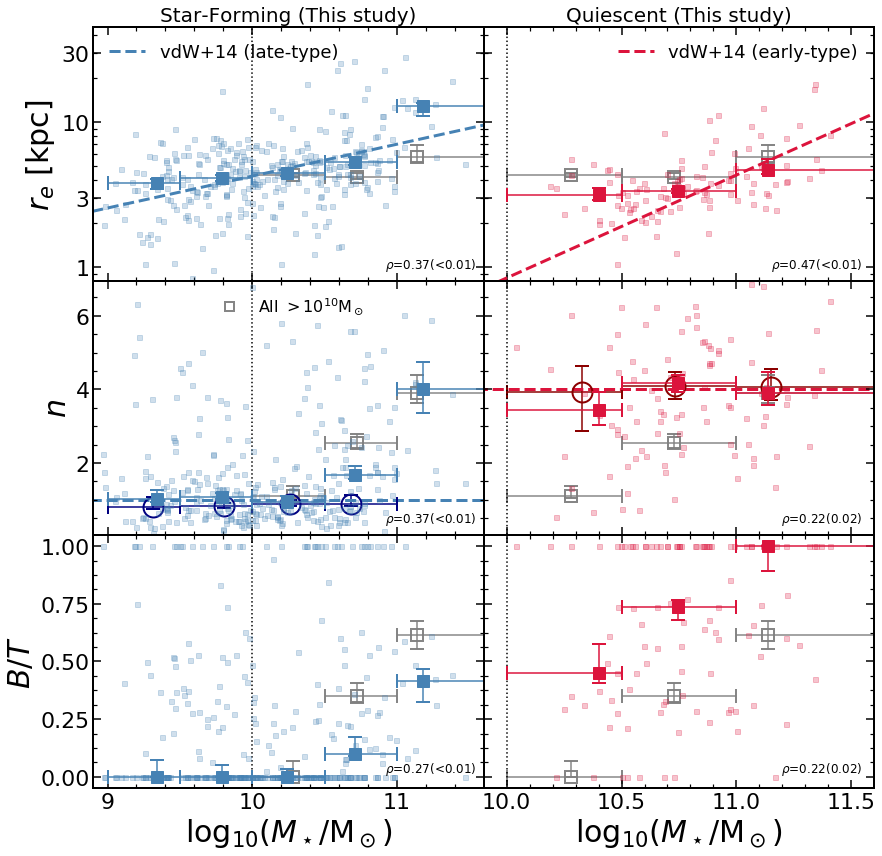}
\caption{\textbf{Top:} Stellar-mass size relation at $z\sim0.84$ for all galaxies in our sample, divided into star-forming (left) and quiescent (right) subsamples. We also show the derived relation for a large sample at similar redshift for star-forming (blue dashed line) and quiescent galaxies (red dotted line) derived by \citet{vanderwel2014}. We find a good agreement between our sample and a magnitude-limited sample at these redshifts, indicating that our sample is representative of the larger population in terms of sizes and stellar masses. \textbf{Middle:} S\'ersic index as a function of stellar mass for galaxies best fit by a single S\'ersic profile. The median value for all galaxies is shown with large squares, and that for the subset of galaxies best fit with one component is shown with large empty circles. We show as horizontal dashed lines the values for an exponential disc (blue, $n=1$) and a classical elliptical (red, $n=4$) profile. The vertical dotted line highlights the stellar mass selection limit of our survey. We find star-forming and quiescent galaxies to align with the classical expectations at lower redshifts, with quiescent galaxies and star-forming galaxies  having profiles typical of ellipticals and typical discs, respectively. \textbf{Bottom:} Bulge-to-total light ratio as a function of stellar mass. The median is shown with large symbols. We find quiescent galaxies to have slightly more prominent bulges than star-forming galaxies at similar stellar masses. We add to all panels the relation for the global sample at $M_\star>$\msun{10} as empty grey squares. In each panel, we show the Spearman correlation coefficient, $\rho$,  and the corresponding probability of an uncorrelated dataset having the same distribution in parentheses (considering only $M_\star>$\msun{10}).}
\label{fig:parametric_mass_sfa}
\end{figure*}

The top panel of Fig. \ref{fig:parametric_mass_sfa}  shows the relation between median galaxy size (measured as the effective radius) as a function of stellar mass for quiescent and star-forming galaxies. We find good agreement with a large sample at similar redshifts \citep{vanderwel2014} as expected given that galaxies in VIS\textsuperscript{3}COS are representative of the larger population at these redshifts. We note that for the quiescent sample, {\citet{vanderwel2014} only fitted the stellar-mass--size relation using galaxies} more massive than $10^{10.3}\mathrm{M_\odot}$. These latter authors also apply a misclassification (possible confusion between star-forming and quiescent galaxies) correction that lowers the weight of large quiescent galaxies and small star-forming galaxies in the joint fit of the stellar mass--size relations. This is likely the reason for the difference between our median value and their best-fit relation at lower stellar masses. Regarding the correlation strength, we find a slightly stronger correlation for quiescent galaxies at $M_\star>$\msun{10} ($\rho=0.47$ when compared to $\rho=0.37$ for star-forming galaxies), but for both populations the correlations are significant, as already found by many studies \citep[e.g.][]{franx2008, vanderwel2014, morishita2014, sweet2017, mowla2019}.

We split each population into three local density bins (see Figs. \ref{fig:parametric_mass_env_SF} and \ref{fig:parametric_mass_env_Q}) to investigate the existence of any dependence of galaxy size on environment at $z\sim1$. For star-forming galaxies, we find no significant difference of the median values with local density and we find weak correlations for each stellar mass subsample. Our results are consistent with those reported by other studies \citep[e.g.][]{lani2013,kelkar2015,tran2017}. Though some studies find differences between field and cluster galaxies \citep[e.g.][, between 7\% and 16\% larger in cluster environments]{cebrian2014,allen2015,allen2016}, these differences are smaller than our error bars and consistent with our results. Regarding the quiescent population we find no significant dependence with environment for galaxies with $10<\log_{10}\left(M_\star/\mathrm{M_\odot}\right)<11$. This is also consistent with results from the literature targeting similar stellar mass ranges \citep[e.g.][]{huertas-company2013,cebrian2014,newman2014,allen2016,saracco2017}. For the most massive quiescent galaxies in our sample, we find larger sizes for galaxies in the highest-local density bin when compared to the two lower-density bins (see Fig. \ref{fig:parametric_mass_env_Q}). It is also the most massive galaxies that have the strongest correlation between size and local density ($\rho=0.27$ and $\sim$17\% probability of being an uncorrelated distribution), despite not being as significant as the correlation found between stellar mass and galaxy size. This is already hinted at in Fig. \ref{fig:parametric_mass_env} and is found in other studies at these high stellar masses \citep[see e.g.][]{papovich2012,saracco2017,yoon2017}.

        \subsection{Prominence of galactic bulges}\label{ssec:bulge_results}

We explore the impact of stellar mass on the steepness of the light profiles in star-forming and quiescent galaxies. We show in Fig. \ref{fig:parametric_mass_sfa} the median one-component S\'ersic index for all galaxies. We find that quiescent galaxies have similar S\'ersic indices, $n\sim4$, at all stellar masses greater than \msun{10}, the typical value for classical ellipticals. For star-forming galaxies, we find a rise of the median value of $n$ with stellar mass, going from $n\sim1$ at \msun{10.25} to $n\sim4$ at $>$\msun{11}. We note, however, that this rise in S\'ersic index can be traced to a change in the structure of star-forming galaxies with stellar mass, from simple discs to disc+bulge systems. As highlighted in Fig. \ref{fig:parametric_mass_sfa}, when considering only those galaxies for which the best fit is a single S\'ersic (where the value of $n$ is the better descriptor of the light profile shape) we find no trends with stellar mass, with the median value of $n$ being the typical value for exponential discs $n\sim1$. We attempt to further split our sample into overdensity bins to explore the impact of environment on galaxy structure and we find no or little difference for samples in different environments (with the exception of a positive trend with environment - $\rho=0.70$ - considering the 11 high-stellar-mass star-forming galaxies populating the two lower-density bins; see Figs. \ref{fig:parametric_mass_env_SF} and \ref{fig:parametric_mass_env_Q}).

We also show in Fig. \ref{fig:parametric_mass_sfa} the median bulge-to-total light ratio ($B/T$) for quiescent and star-forming galaxies in different stellar mass bins. Overall we find quiescent galaxies to have higher values of $B/T$ than star-forming galaxies at stellar masses greater than \msun{10}, which is expected given the more bulge-dominated nature of quiescent galaxies \citep[e.g.][]{wuyts2011,keunho2018,morselli2018}. Regarding the trend with stellar mass, we find that both quiescent and star-forming systems show an increase of $B/T$ with increasing stellar mass (weak, non-negligible correlation - $<1$\% - likely of being an uncorrelated distribution). In the quiescent population  $B/T$  rises from $\sim0.4$ to $\sim1$ (from $\sim$\msun{10.25} to $\sim$\msun{11.25}) while for the star-forming population it rises from $\sim0$ to $\sim0.4$ in the same stellar mass interval. In Figs. \ref{fig:parametric_mass_env_SF} and \ref{fig:parametric_mass_env_Q} we show the dependence of $B/T$ on environment for the shown stellar mass bins for both populations. In the case of star-forming galaxies, there is no significant trend with local density. For quiescent galaxies, we might see a hint of a trend when considering the median values for galaxies more massive than \msun{10.5}, but the correlations are very weak.

        \subsection{Morphology trends with a model-independent approach}\label{ssection:npar_results}

\begin{figure*}
\centering
\includegraphics[width=0.85\linewidth]{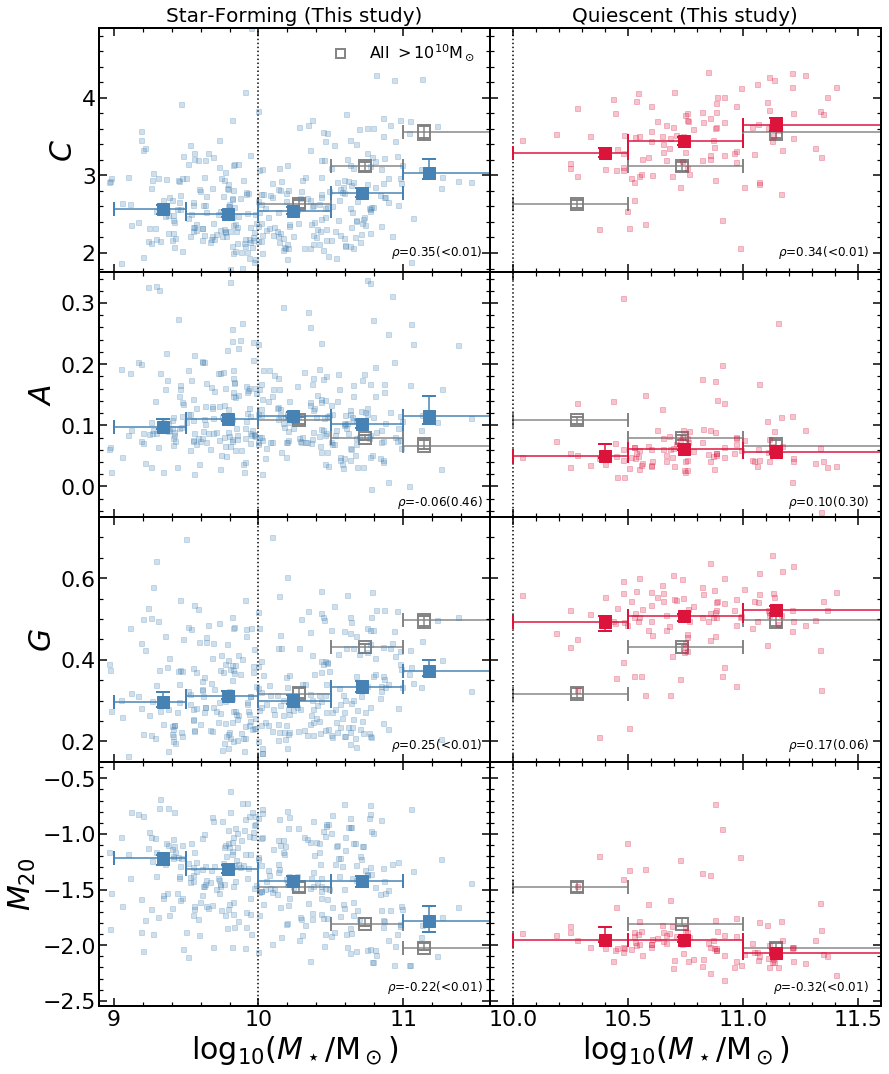}
\caption{Light concentration (top), image asymmetry (middle top), Gini coefficient (middle bottom), and moment of light (bottom) as a function of stellar mass. We add to all panels the relation for the global sample at $M_\star>$\msun{10} as empty grey squares. In each panel, we show the Spearman correlation coefficient, $\rho$,  and the corresponding probability of an uncorrelated dataset having the same distribution in parenthesis (considering only $M_\star>$\msun{10}). We find quiescent galaxies to have a higher concentration of light than star-forming galaxies at similar stellar masses. We also find quiescent galaxies to have less disturbed profiles at stellar masses greater than $10^{10}\mathrm{M_\odot}$. This is likely a reflection of the lack of star formation that is clumpier in nature \citep[][]{conselice2003}. We also find quiescent galaxies to have their light concentrated on a smaller area (higher value of $G$) than star-forming galaxies at similar stellar masses. Finally, quiescent galaxies are smoother (lower values of $M_{20}$) at all stellar masses, as also seen in the asymmetry parameter.}
\label{fig:nonparametric_mass_sfa}
\end{figure*}

As detailed in Section \ref{ssec:nonpar}, there are a number of quantities that can describe the light profiles of galaxies without the assumption of a physical model. In Fig. \ref{fig:nonparametric_mass_sfa}, we present the properties of star-forming and quiescent galaxies as a function of stellar mass. We find that quiescent galaxies have higher concentration indices than star-forming galaxies at all stellar masses. We also find that  galaxies with higher stellar mass (from $\sim10^{10.25}\mathrm{M_\odot}$ to $\sim10^{11.25}\mathrm{M_\odot}$) have higher concentration values in both populations (correlations with stellar mass are equally strong for both populations). In quiescent galaxies the median value of the light concentration ($C$) rises from $3.29\pm0.05$ up to $3.64\pm0.09$ and in star-forming galaxies it rises from $2.54\pm0.05$ to $3.0\pm0.1$. The fact that quiescent galaxies have higher concentration values than their star-forming counterparts is consistent with them having elliptical or bulge-dominated morphologies. 

We also show in Fig. \ref{fig:nonparametric_mass_sfa} the median asymmetry of galaxy light profiles. We find that neither star-forming nor quiescent galaxies' asymmetry is correlated with their stellar mass (low correlation coefficient, no significant change in the median values). Considering galaxies above our stellar mass selection limit ($10^{10}\mathrm{M_\odot}$), we find quiescent galaxies to have lower asymmetry ($A\sim0.05-0.06$) than star-forming galaxies ($A\sim0.10-0.12$) at all stellar masses. This difference in asymmetry indicates that quiescent galaxies have smoother light profiles when compared to star-forming galaxies, which have a clumpier light profile due to blue star-forming clumps.

In Fig. \ref{fig:nonparametric_mass_sfa} we show the results of a different set of morphology diagnostics. We find that quiescent galaxies have a higher percentage of their light concentrated on a smaller area (higher Gini coefficient - $G$) when compared to star-forming galaxies at similar stellar masses. Considering galaxies with stellar masses above $10^{10}\mathrm{M_\odot}$, we find a negligible increase in the median value of $G$ for quiescent galaxies (from $0.49\pm0.02$ to $0.52\pm0.01$) and a steeper increase for star-forming galaxies (from $0.30\pm0.01$ to $0.37\pm0.02$). The correlation coefficient also points to a stronger trend for star-forming galaxies, despite the large scatter. When considering the value of the moment of light ($M_{20}$), which measures the concentration of the brightest regions and is sensitive to the existence of multiple clumps, we find a global trend for galaxies with high stellar mass to have lower values of $M_{20}$ (higher concentration of the brightest regions, irrespective of clumpy substructures). We find non-negligible but weak correlations with stellar mass for both populations. We also find quiescent galaxies to have higher flux concentration when compared to star-forming galaxies of similar stellar masses. The combination of these two quantities highlights the difference between quiescent galaxies having a higher concentration of their flux and being less likely to have clumpy substructures when contrasted to their star-forming counterparts. 

We split both populations into different bins of local density and find no statistically significant differences among different environments at fixed stellar mass bins for each population in all the presented tracers (see Figs. \ref{fig:nonparametric_mass_env_SF} and \ref{fig:nonparametric_mass_env_Q}). The trends reported in this section all hint at quiescent galaxies having morphologies characteristic of elliptical (or bulge-dominated) light profiles whereas star-forming galaxies resemble more exponential discs with a larger degree of clumpiness or asymmetry in their light profiles.

        \subsection{Local density impact on visual morphology}

We defined in Section \ref{ssection:visual_class} three different morphological classes based on Galaxy Zoo classifications of HST data. In this section, we explore the impact of local density on the fraction of galaxies for each of the defined classes: ellipticals, discs, and irregulars. We restrict our analysis to galaxies more massive than  $10^{10}\mathrm{M_\odot}$  (our selection limit).

\begin{figure}
\includegraphics[width=\linewidth]{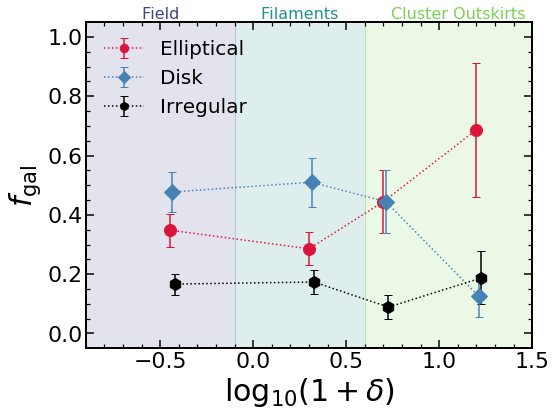}
\caption{Fraction of galaxies more massive than $10^{10}\mathrm{M_\odot}$ of a given galaxy morphology (see Section \ref{ssection:visual_class}) as a function of local density. Errors on the fractions are computed from Poisson statistics. We show as coloured vertical regions the likely association between local density and density regions. We note that we find no significant differences between field-like and filament-like densities. We do find a rise in the fraction of ellipticals and a decline of disc-like morphologies towards the densest regions probed here.}
\label{fig:density_morph}
\end{figure}

Figure \ref{fig:density_morph} reveals the differences in the fraction of observed morphologies for all massive galaxies in our sample. At lower densities (field- and filament-like regions) we find fractions of disc galaxies to be similar ($48\pm6$\% and $51\pm8$\%, respectively). The same scenario applies to elliptical galaxies ($34\pm6$\% and $28\pm6$\%, respectively) and irregular galaxies ($17\pm4$\% in both local density bins). As we move towards higher-density regions, we find an increase in the fraction of elliptical galaxies (up to $69\pm23$\%)  and a strong decline in the fraction of disc galaxies (down to $13\pm7$\%). For irregular galaxies, there is a small drop to $9\pm4$\% and then rise to $19\pm9$\% towards the highest densities, but our values are consistent with a constant fraction at all local densities. This result hints at an established morphology--density relation at $z\sim0.84$.

\begin{figure}
\includegraphics[width=\linewidth]{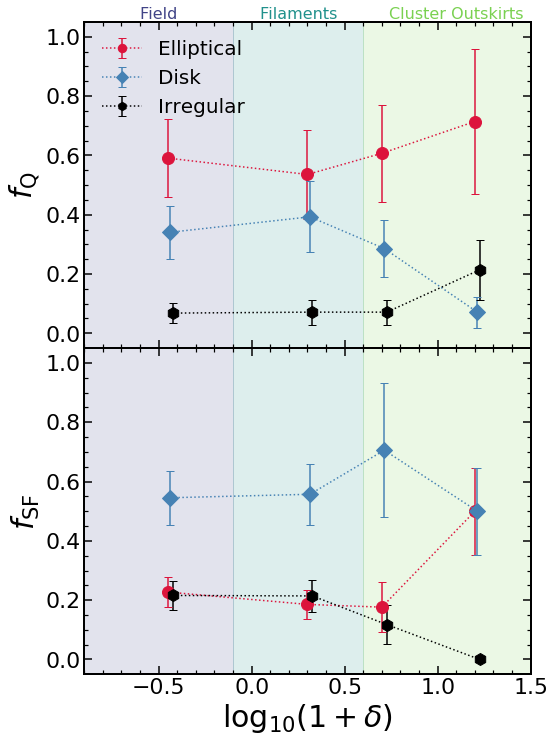}
\caption{Fraction of galaxies more massive than $10^{10}\mathrm{M_\odot}$ of a given galaxy morphology (see Section \ref{ssection:visual_class}) as a function of local density for quiescent (top) and star-forming (bottom) galaxies. Errors on the fractions are computed from Poisson statistics.  We show as coloured vertical regions the likely association between local density and density regions. For quiescent galaxies, we see a nearly constant fraction of elliptical galaxies with density and an increase in irregular morphologies at the expense of disc galaxies in the densest regions. For star-forming galaxies, we might find a small trend in the densest bin with an increase in ellipticals and a decline in irregular galaxies (and also potentially discs, though the disc fraction is compatible with no environmental influence).}
\label{fig:density_morph_sfa}
\end{figure}

We note, however, as discussed in Sections. \ref{ssection:sizes}-\ref{ssection:npar_results}, that if we split our sample into star-forming and quiescent populations, we find little effect of local environment on quantitative morphology within each population. We also explore here the fraction of each class for these two populations in Fig. \ref{fig:density_morph_sfa}. For quiescent galaxies, we find that the fraction of ellipticals dominates at all environments, and we observe no change with local density (nearly constant fraction at $\sim$60\%). For quiescent galaxies with disc morphology, we find a constant fraction in the field- and filament-like densities ($\sim$35\%) and then a drop towards higher densities (down to $7\pm5$\%). We find that quiescent galaxies with irregular morphologies make up $\sim$7\% in lower-density regions and then rise to $21\pm10$\% in the highest density bin, surpassing the fraction of discs at these densities, indicating an increase in galaxy interactions at the higher densities.

For star-forming galaxies, we find disc morphologies to be the most common class at all densities ($\sim$57\%) with little change across different densities. For star-forming ellipticals, we find a nearly constant fraction for the three lower-density bins (at $\sim$30\%) and then rise to $50\pm15$\% at the highest-density bin. We also find a decrease in the fraction of irregular star-forming galaxies from the field- and filament-like regions ($\sim$20\%) down to 0\% at the highest-density bin in the sample. 

Our results hint at an effect of local density on galaxy morphology. In regards to the quiescent population, we see a trend of change from red disc galaxies to irregular galaxies, likely related to the tidal disruption of galactic discs by interactions with other cluster members. For the star-forming population, we see a change of disc and irregular galaxies into elliptical galaxies, likely through mergers \citep[e.g.][]{bournaud2007,kormendy2009,taranu2013,martin2018}.

%%%%%%%%%%%%%%%%%%%%%
%%%%%%%% SECTION 6 %%%%%%
%%%%%%%%%%%%%%%%%%%%%

\section{Discussion}\label{section:discussion}

We study galaxy morphology on a sample of approximately 500 spectroscopically confirmed galaxies in and around a superstructure in COSMOS at $z\sim0.84$. Although we find that the morphological measurements of a galaxy depend on both its environment and its stellar mass, when we split the sample into star-forming and quiescent systems, such morphological trends weaken significantly or vanish completely. In the following, we try to explain this with a simple model.

        \subsection{Structural dependence predicted from the quiescent fraction}\label{ssection:model_fq}

\begin{figure*}
\includegraphics[width=\linewidth]{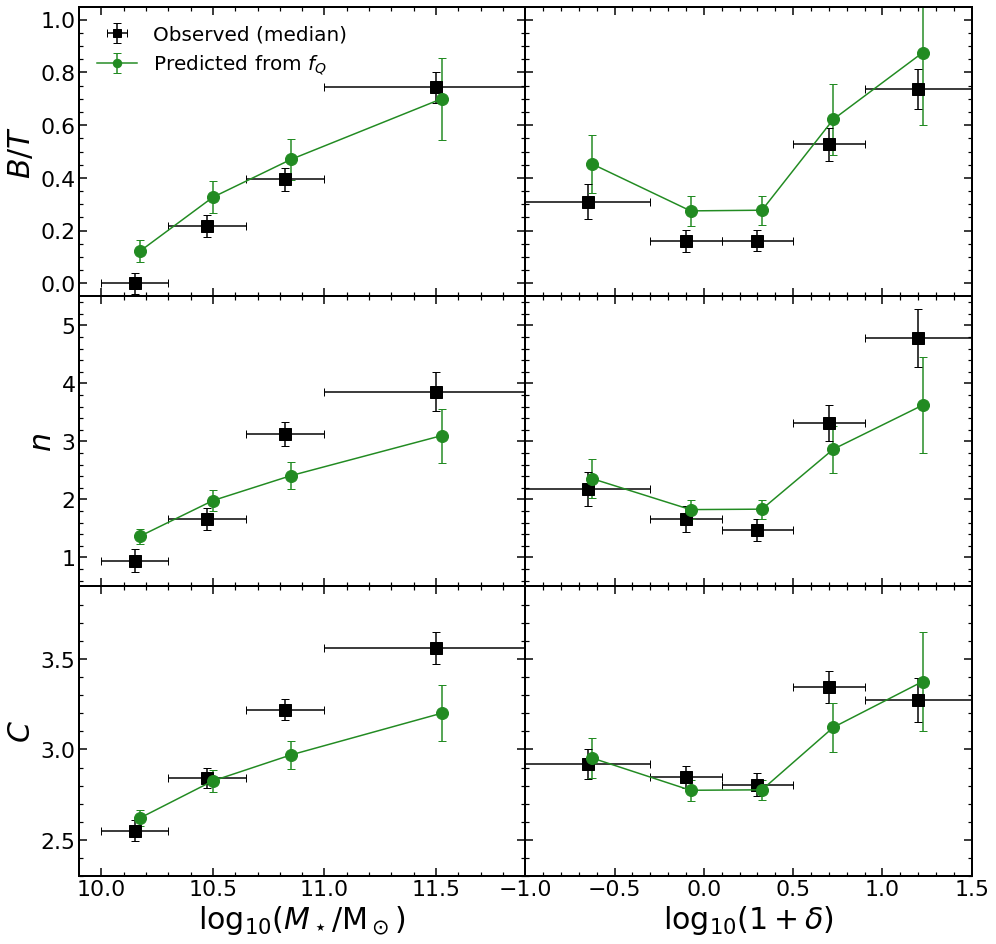}
\caption{Predicted (green circles) and observed median (black squares) values of the bulge-to-total light ratio (top), S\'ersic index (middle), and light concentration (bottom) for galaxies more massive than \msun{10}. The predicted values are based on a simple model (see Section \ref{ssection:model_fq}) that predicts stellar mass or environmental dependence of any property based on the fractions of quiescent and star-forming galaxies at different stellar masses and in different environments. We find a good agreement between predicted and observed values, indicating that the perceived effects of galaxy structure and morphology on stellar mass and environment are tightly correlated with the fraction of star-forming and quiescent populations in each bin.}
\label{fig:model_fq}
\end{figure*} 

In Section \ref{section:results_sfr} we find that there is a small dependence of structural measurements on stellar mass for galaxies split into star-forming and quiescent (see e.g. S\'ersic index in Fig. \ref{fig:parametric_mass_sfa}). We also find little or no dependence of morphological indicators (both quantitative and qualitative) on local density when we split the sample into star-forming and quiescent systems (see Figs. \ref{fig:parametric_mass_env_SF} through \ref{fig:nonparametric_mass_env_Q}). We also show (see e.g. Figs. \ref{fig:parametric_mass_env}, \ref{fig:nonparametric_mass_env}, and \ref{fig:density_morph}) that we find structural and morphological dependence on stellar mass and environment when considering the global sample at stellar masses $>$\msun{10}. We attempt here to explain the observed changes with density as a consequence of the change in the fraction of each population (star-forming or quiescent) that is present at each environment and stellar mass bin. 

\citet[][]{paulino-afonso2018} show the dependence of the quiescent fraction on stellar mass and environment for galaxies more massive than $10^{10}\mathrm{M_\odot}$, and find that it strongly increases with stellar mass and also from intermediate- to high-density regions \citep[see also e.g.][]{peng2010b,cucciati2010b,muzzin2012,darvish2014,darvish2016,darvish2018,hahn2015}. To test our assumption we assume that the average property $x$ in a given stellar mass or environment bin is a combination of the individual properties of each population weighed by its fraction in that bin.  We can then parametrize the dependence of $x$ on stellar mass or environment as a function of the fraction of quiescent galaxies $f_Q$ on the binned quantity:

\begin{equation}
x = \frac{x_\mathrm{SF}N_\mathrm{SF} + x_\mathrm{Q}N_\mathrm{Q}}{N_\mathrm{T}} = x_\mathrm{SF}(1-f_Q) + x_\mathrm{Q}f_Q.
\label{eq:fq_model}
\end{equation}

This can subsequently be used to predict the expected values of any property if the fraction of quiescent objects is the driving influence of the observed dependences. For example, we can derive the median $B/T$ as a function of stellar mass or environment, assuming that all star-forming galaxies have $B/T=0$ (exponential discs) and all quiescent galaxies have $B/T=1$ (classical ellipticals). To compute the median properties from our observations, we also assign a value of $B/T$ for galaxies best fit with a single S\'ersic profile ($B/T=0$ if $n<2.5$ and $B/T=1$ if $n>2.5$; see Section \ref{ssection:par_results}). We show in the bottom panel of Fig. \ref{fig:model_fq} the resulting prediction compared to the median observed values of $B/T$.
We can apply this method to other quantities, and we highlight the light profile shape traced by parametric (S\'ersic index $n$) and non-parametric (light concentration $C$) quantifications of galaxy structure in Fig. \ref{fig:model_fq}. For the case of $n$ we use a constant value of $n_\mathrm{SF} = 1$ and $n_\mathrm{Q}=4$ (based on single S\'ersic best-fit relations illustrated in Fig. \ref{fig:parametric_mass_sfa}). Considering the median observed value of $n$ per bin of stellar mass, we can broadly reproduce the trend; although the trend with stellar mass is steeper (stronger variation, meaning a stronger underlying correlation with stellar mass for the sub-populations) than what is predicted from the quiescent fraction. In terms of the dependence on environment, we find remarkably good agreement between the two independent quantities (median $n$ and $f_Q$). We find a similar result when considering the model-independent light concentration $C$ as the morphology tracer (using a simple constant value of $C_\mathrm{SF} = 2.5$ and $C_\mathrm{Q}=3.5$ in equation \ref{eq:fq_model}).

Since we assume the most straightforward dependence of structural parameters on stellar mass and environment for each population, that is, a constant value, it is natural that the discrepancy between the predicted and observed values is larger when our assumption of constancy is farther from the truth. Moreover, since the correlations of the studied parameters with local density are the weakest of the two, we find that the match between our predictions and the observed median values is better in this case.
The good agreement between the observed and the predicted value from our straightforward model is a strong argument in favour of the morphology--density relation being tightly correlated with the fractions of quiescent and star-forming galaxies in different environments \citep[e.g.][]{calvi2018}. This scenario is also consistent with the strongest impact of environment appearing to take place on the quiescent fraction \citep[e.g.][]{darvish2014,darvish2016,darvish2018}. Furthermore, this would also mean that processes that affect galaxy morphology, either in the formation of galaxies or posterior interactions, might also impact star formation \citep[e.g.][]{martig2009, wuyts2011}, although they might happen at different stages in their evolution \citep[e.g.][]{bundy2010}.

It is possible that the growth of a bulge is induced by a higher rate of interactions in higher-density environments since several studies point to major and minor mergers as mechanisms for bulge growth \citep[e.g.][]{eliche-moral2006,hopkins2010, querejeta2015, brooks2016}. In the local Universe, merger-induced star formation is important \citep[e.g.][]{lambas2012,ellison2013,scudder2015} and can play a role in the change of not only galaxy colour \citep[see also e.g.][]{ellison2018}, but also structure required to explain the observations in our study. In this scenario, the bulge prominence is correlated with the probability of the galaxy being quenched, with the quenched fraction being higher for high-$B/T$ systems \citep[see e.g.][]{lang2014}. A natural consequence of this is that the  $B/T$ (and also more generally the S\'ersic index and light concentration which measures similar properties) ratio of galaxies is correlated with $f_Q$, as we show in Fig. \ref{fig:model_fq} \citep[see also][]{keunho2018}.

        \subsection{Morphology--density relation at $z\sim0.84$}

Some studies show evidence for a correlation between morphology and environment up to $z\sim1$ \citep[e.g.][]{tasca2009}. We find that such a relation is also present in our sample (see Fig. \ref{fig:density_morph}). However, we also show that the impact of local density on galaxy structure among blue star-forming and red quiescent galaxies is negligible. What we find is consistent with the fractions of red and blue galaxies changing with environment, and morphology tracing that change as well (see Fig. \ref{fig:model_fq}). This again suggests, as discussed before, that the environment is mostly correlated with the quenched fraction, and does not affect the morphology of star-forming or quiescent galaxies at $z\sim1$. 

The differences in galaxy morphology for quiescent and star-forming galaxies have long been studied and established up to $z\sim1$ \citep[e.g.][]{strateva2001, bamford2009, mignoli2009, wuyts2011, whitaker2015, krywult2017}. Other studies show that the environmental dependence of galaxy morphology is tightly correlated with colour \citep{poggianti2008,skibba2009, bait2017}. This is in agreement with our findings that when splitting our sample for star formation activity, the dependence on the environment is small \citep[see also e.g.][]{papovich2012,huertas-company2013b, huertas-company2013, lani2013,cebrian2014, newman2014, kelkar2015, allen2015, allen2016, saracco2017}. The existence of such a correlation hints at a coherent transformation both in star formation and morphology for galaxies in different environments. This has already been seen in some studies targeting green valley galaxies (with colours between the red sequence and the blue cloud) where morphologies between exponential discs and classical ellipticals are found \citep[e.g.][]{mendez2011,coenda2018,gu2018}. However, a difference in colour does not always translate to a difference in morphology for these sources \citep[e.g.][]{schawinski2014, vulcani2015} and both internal and external processes are required to explain such evolution across the green valley \citep[e.g.][]{mahoro2017, kelvin2018, nogueira2018}.

The local morphology--density relation has a category of galaxies that plays a pivotal role in the observed trends but is not included in our analysis of visual morphology, namely  S0 galaxies \citep[e.g.][]{dressler1980,dressler1984}. However, these are less common at higher redshifts \citep[$z\sim0.5-0.8$ e.g.][]{dressler1997, desai2007, poggianti2009b, just2010}. Given the existing classifications, S0 galaxies can fall into either the disc or elliptical categories, depending on the inclination with respect to the line of sight. Edge-on S0s are more likely to be classified as discs, while face-on S0s can be mistaken for ellipticals using our scheme. This means that we are not exploring the full scenario of morphological transformation in dense environments, but rather a simplified version of this correlation, considering only the two major classes of the original \citet[][]{hubble1926} classification scheme (spiral discs and ellipticals).  A more refined classification scheme would require a specific classification scheme with an identifiable option for S0 galaxies and a larger sample to be able to statistically disentangle the larger number of classes we would have to deal with, but this is out of the scope of this manuscript.

%%%%%%%%%%%%%%%%%%%%%
%%%%%%%% SECTION 7 %%%%%%
%%%%%%%%%%%%%%%%%%%%%

\section{Conclusions}\label{section:conclusions}

We study the influence of stellar mass and environment on galaxy morphology with the VIS\textsuperscript{3}COS survey in and around a superstructure at $z\sim0.84$ in the COSMOS field. We present our results on the bulge-to-disc decomposition of light profiles, non-parametric morphology, and visual classification. We also  separately study star-forming and quiescent galaxies selected in the NUV-r-J colour space. Our results can be summarised as follows.

\begin{itemize}
\setlength\itemsep{0.5em}
\item There is an environmental dependence of S\'ersic indix and $B/T$ in different stellar mass bins when considering the entire sample, with denser environments having galaxies with higher S\'ersic indices and $B/T$ for fixed stellar mass.
\item We find that stellar mass is a stronger predictor of galaxy structure and morphology (stronger correlations) than local density for all quantities studied here.
\item We find that for galaxies more massive than \msun{11} there is an increase in size ($\sim40\%$) from low- and intermediate-density regions to high-density regions. Less massive (between \msun{10} and \msun{11}) galaxies show no dependence on local density.
\item Quiescent galaxies are smaller than their star-forming counterparts at similar stellar masses. We find no difference between different environments for star-forming galaxies. For quiescent galaxies, we see a change in galaxy size from low- and intermediate- to high-density regions in the most massive bin ($>$\msun{11}), which drives the differences found when looking at the full sample.
\item Galaxies best fit with a single profile show a clear morphology--colour dichotomy. Quiescent galaxies have median S\'ersic indices comparable to classical ellipticals ($n\sim4$), and star-forming galaxies show profiles close to exponential discs ($n\sim1$).
\item We also find differences in light profiles with non-parametric morphology. Quiescent galaxies have smoother profiles (lower asymmetry and $M_{20}$) and have more concentrated light profiles (higher concentration and Gini coefficient) than star-forming galaxies.
\item We find evidence for the existence of a morphology--density relation at $z\sim0.84$ when looking at the sample as a whole, but this is  less pronounced when splitting into star-forming and quiescent subsamples.
\item When combined, our results point to a tight correlation between morphology and colour, with quiescent and star-forming galaxies showing little dependence on environment. We can reproduce the observed trends of structure and morphology (traced by $B/T$, $n$, and $C$) with a local density as a natural consequence of the change in the quenched fraction for different environments. 
\end{itemize}

We thus find that environmental dependences of galaxy structure and morphology exist when considering the entire sample. However, those dependences are much less pronounced when considering only the star-forming or the quiescent subsamples. Based on our results, we argue that colour as well as structure and morphology are affected by environment, and this is manifested through a varying fraction of blue discs to red ellipticals from low- to high-density regions. Such a tight correlation between star formation and morphology implies that physical mechanisms responsible for regulating star formation must also act in changing the structure and morphology of galaxies, such as galaxy mergers or strong feedback events. The subtle effects of both the stellar mass and environment allow better constraints on the possible scenarios for galaxy evolution across different environments. A better sampling of galaxies in the transition phase (in filaments and/or in the green valley) is necessary to unequivocally pinpoint the mechanisms responsible for the observed changes with stellar mass and environment.

%%%%%%%%%%%%%%%%%%%%%
%%%%% ACKNOWLEDGEMENTS %%%
%%%%%%%%%%%%%%%%%%%%%

\begin{acknowledgements}
We thank the anonymous referee for the insightful and useful comments that helped improve the quality and clarity of the manuscript. This work was supported by Funda\c{c}\~{a}o para a Ci\^{e}ncia e a Tecnologia (FCT) through the research grant UID/FIS/04434/2013. APA is a PhD::SPACE fellow who acknowledges support from the FCT through the fellowship PD/BD/52706/2014. DS acknowledges financial support from Lancaster University through an Early Career Internal Grant A100679. BD acknowledges financial support from NASA through the Astrophysics Data Analysis Program (ADAP), grant number NNX12AE20G, and the National Science Foundation, grant number 1716907.

This work was only possible by the use of the following \textsc{python} packages: NumPy \& SciPy \citep{walt2011,jones2001}, Matplotlib \citep{hunter2007}, and Astropy \citep{robitaille2013}.
\end{acknowledgements}

%%%%%%%%%%%%%%%%%%%%%
%%%%%% REFERENCES %%%%%%%
%%%%%%%%%%%%%%%%%%%%%

\bibliographystyle{aa}
\bibliography{refs}

%%%%%%%%%%%%%%%%%%%%%
%%%%%% APPENDICES %%%%%%%
%%%%%%%%%%%%%%%%%%%%%

\begin{appendix}

\section{Technical details on morphological parameters}

        \subsection{Statistical choice of best-fit parametric model}\label{app:bic}

One is free to choose a model with as many components as one wants to fit every galaxy. However, to get physically meaningful results from fitting galaxy images, one should take caution with over fitting the data by choosing models that are too complex when compared to what is needed to fit the actual data. There have been some statistical criteria to decide whether or not a complex model should be used \citep[e.g.][]{simard2011,kelvin2012proc,meert2013,bruce2014a,margalef-bentabol2016}. The BIC, used for example by \citet{kelvin2012proc} and \citet{bruce2014a}, is a measure of how good a model fits the data one wants to describe. In the case of nested models, it penalizes those with a higher number of free parameters. The model is described by
\begin{equation}
BIC = \chi^2 +k\ln(N),
\end{equation}
where $\chi^2$ is the measure of the global goodness of the fit given by \textsc{GALFIT}, $k$ is the number of free parameters of the model we are considering and $N$ is the number of contributing data points to the analysis of the model that is taken to be the area, in pixels, of the object being considered. Given two models we can compute the difference in this estimator with
\begin{equation}
\Delta BIC = BIC_c - BIC_s = (\chi^2_c-\chi^2_s) +(k_c-k_s)\ln (N),
\end{equation}
where $s$ and $c$ denote the simple (one profile) and complex (bulge+disc) models, respectively. The preferred model is the one with the lowest BIC value. In a strict sense, if $\Delta BIC<0$ then the complex model is to be chosen over the simplest one. However, to be sure that the complex model is more than simply marginally better than a single profile, we apply a stricter rule for which $\Delta BIC< -10$ \citep[e.g.][]{kelvin2012proc}.

        \subsection{Non-parametric computation}\label{app:np}

                \subsubsection{Light concentration}

The concentration index $C$ is defined as the ratio of the 80\% to the 20\% curve of growth radii within 1.5 times the \citet[][\!\!, $r_p$]{petrosian1976} radius for a parameter $\eta=F(r)/\int_0^{r}F(r)=0.2$ \citep[see e.g.][]{bershady2000}. With that radius we compute the flux using elliptical apertures centred on the light-weighted centre of the galaxy up to which 20\% and 80\% of the light is contained. We then compute $C$ via
\begin{equation}
C = 5\log\left(\frac{r_{80}}{r_{20}}\right).
\end{equation}

This parameter allows one to separate between concentrated objects such as ellipticals and more extended sources such as spirals or irregulars. Using this definition the values of $C$ range from about 2 to 5, where $C > 4$ usually indicates spheroid-like systems, $3<C<4$ indicates disc galaxies and the lower values of $C$ are from low surface brightness objects or sometimes from multi-component systems \citep[see e.g.][]{conselice2003}.

                \subsubsection{Asymmetry}

The asymmetry index $A$ measures the strength of non-axis-symmetric features of an image $I$ by comparing it to a  version of itself rotated by 180 degrees, $I^{180}$. Since we expect asymmetric features on irregular galaxies usually associated with galaxy--galaxy interactions, this index is very useful to identify ongoing galaxy mergers. It also correlates with ongoing star formation as individual star-forming regions in a larger galaxy can also produce asymmetric flux distributions \citep{bershady2000,conselice2000,conselice2003}. We compute the index $A$ as
\begin{equation}
A = \frac{\sum_{i,j} | I_{i,j} - I_{i,j}^{180}| }{\sum_{i,j} I_{i,j}} - B_{180},
\end{equation}
where $I_{i,j}$ is the intensity at the pixel (i,j) and $B_{180}$ is the intensity of the background asymmetries. The centre around which the image is rotated is an important parameter, and there are difficulties in having a well-defined galaxy centre. We follow the method of \citet{conselice2000} and iterate the centre position following a gradient-step approach starting from the light-weighted centre to find the local minimum of $A$ within the segmentation map. To compute $B_{180}$, we use the median of 100 different sky patches of the same size of the image on which we compute $A$ and extract from regions around the object of interest.

                \subsubsection{Gini coefficient}

The Gini coefficient, $G$, measures the concentration of light within the pixels belonging to the  segmentation map of the galaxy. There is some correlation between $G$ and $C$ simply because more-concentrated galaxies tend to have their light distributed over a small number of pixels, therefore leading to high values of $G$ and $C$. Reversely, low and shallow surface brightness profiles tend to have their light more equally distributed, leading to lower values of $G$ and $C$. However, the Gini coefficient will differ from the Concentration parameter in those cases where there is a concentration of high-flux pixels away from the projected centre of the galaxy (e.g. multi-clump galaxy). This index is derived from the Lorenz curve that is a rank-ordered cumulative distribution function of the  pixel values of a galaxy:
\begin{equation}
L(p) = \frac{1}{\bar X}\int_0^p F^{-1}(u)du,
\end{equation}
where $F(u)$ is the cumulative distribution function, $p$ is the percentage of the fainter pixels normalized, and $\bar X$ is the mean pixel flux. The Gini coefficient is then defined as the ratio of the curve $L(p)$ to the equality curve $L(p)=p$. In a discrete population, it can be computed as
\begin{equation}
G = \frac{1}{2\bar X n(n-1)}\sum_{i,j}^n |X_i-X_j|,
\end{equation}
where $n$ is the number of pixels of the galaxy. $G=0$ if all the pixels have the same nonzero flux and $G=1$ if all the flux is contained in one pixel. An efficient way to compute this coefficient is to first sort the pixels of the galaxy in increasing order of flux and then simply compute
\begin{equation}
G = \frac{1}{\bar X n(n-1)}\sum_i^n (2i-n-1)X_i.
\label{gini3}
\end{equation}

Since this coefficient takes into account all pixels of the object, it is very sensitive to the segmentation map associated with the galaxy \citep[see][]{lotz2004}. The inclusion of background flux will increase the value of $G,$ while not taking into account low surface brightness pixels will decrease its value. We note that direct comparison to other results in the literature needs to be done with caution as different definitions of the segmentation map can yield different Gini values for the same galaxies. While this affects the absolute value of $G$, any relative comparison within our sample is valid since it is all computed using the same definition for the segmentation map.  

                \subsubsection{Moment of light}

The index $M_{20}$ is also a measure of light concentration. However, being independent of a specific definition of centre or on having elliptical/circular apertures is less sensitive to asymmetries in the light profile. It is thus a more robust measure for galaxies with multiple bright clumps within a single segmentation map. The total second-order moment $M_{tot}$ is computed by summing the flux in each pixel $I_{i}$ multiplied by the squared distance to the centre of the galaxy. In this case, the centre of the galaxy is that which minimizes $M_{tot}$:
\begin{equation}
M_{tot} = \sum_i^n I_i [(x_i-x_c)^2+(y_i-y_c)^2].
\end{equation}

The index $M_{20}$ is then the normalized sum of the brightest 20\% of pixel values taken from a list of intensity sorted values in a descending order:
\begin{equation}
M_{20} = \log_{10} \left( \frac{\sum_i^N I_i [(x_i-x_c)^2+(y_i-y_c)^2]}{M_{tot}} \right)
,\end{equation}
with the sum considering the pixels that obey $\sum I_i<0.2I_{tot}$ where $I_{tot}$ is the total flux of the galaxy inside the segmentation map region. We normalize by $M_{tot}$ so that this parameter is independent of either total flux or galaxy size. 

        \subsection{Visual classification selection}\label{app:visualclass}

To create subsets of different morphological types \citep[elliptical, disc, or irregular, see e.g.][]{paulino-afonso2018b} we use mainly the results from the first and second tiers \citep[][\!\!, Figure 4]{willett2017}. In a brief explanation, the user is presented with an image and is asked to answer a set of pre-defined questions. The first question is to categorize the galaxy into one of three categories: smooth, features, or star/artefact. If a smooth morphology is chosen, the user is then asked to classify the shape into completely round, in between, or cigar-shaped. If on the other hand, the user classifies the galaxies as having features, then it should classify the galaxy as being clumpy or not. Usually, disc galaxies are classified as non-clumpy featured galaxies. At the end of the process, all users are asked if they find anything to be anomalous (e.g. rings, tails, asymmetries, mergers, disturbed galaxies) which can be used to identify irregular galaxies. The final results for each galaxy are given as the fraction of users that have answered each given possibility.

\section{Mass dependence for the global sample}

\begin{figure*}
\centering
\includegraphics[width=\linewidth]{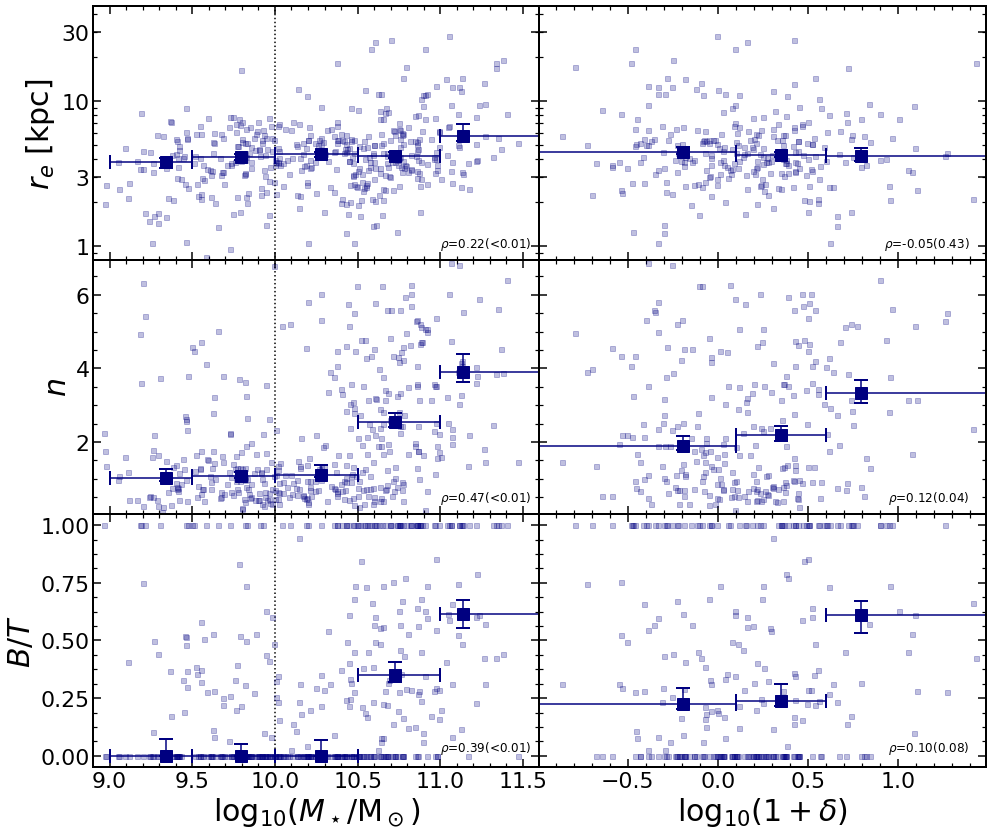}
\caption{Dependence of $z\sim0.84$ galaxy sizes (top), S\'ersic indices (middle), and the bulge-to-total ratio (bottom) on the stellar mass (left) and the environment (right)  for the global sample. In each panel, we show the Spearman correlation coefficient, $\rho$, and the corresponding probability of an uncorrelated dataset having the same distribution in parentheses (the coefficient is computed for galaxies with $M_\star>$\msun{10}).}
\label{fig:parametric_mass_all}
\end{figure*}

We explore in this appendix the global correlations of the studied parameters with stellar mass and local density. We show in Fig. \ref{fig:parametric_mass_all} the dependence of the parametric quantities studied in this manuscript as a function of stellar mass and environment for the global sample. For each parameter, we compute the Spearman correlation coefficient and the corresponding probability of the observed distribution being random for galaxies more massive than \msun{10}. We find that correlations between structural parameters and stellar mass are always stronger (less likely to be random) than that found for correlations with local density. We also find that the luminosity profile shape (traced by $n$ and $B/T$) correlates more strongly with stellar mass and local density than the galaxy size does, the latter being roughly constant for varying stellar mass and local density.

\begin{figure*}
\centering
\includegraphics[width=\linewidth]{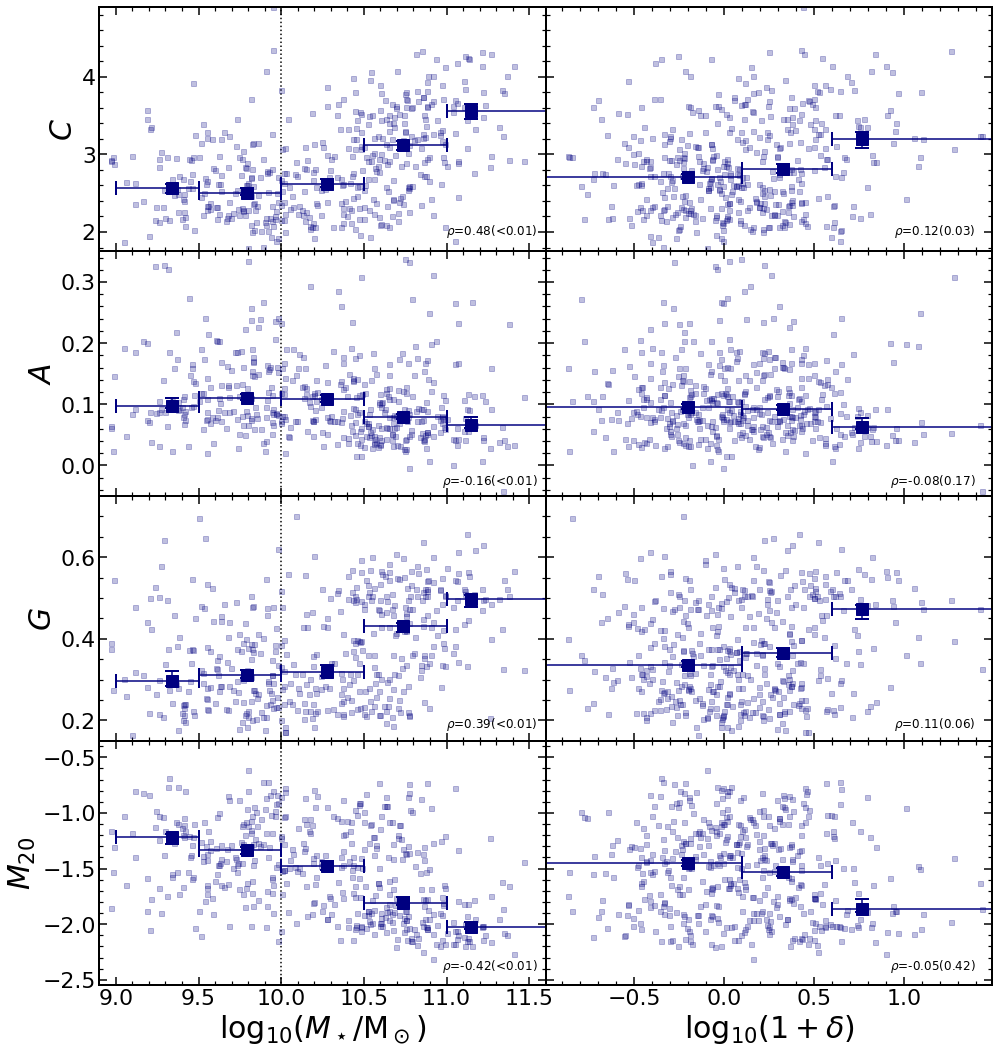}
\caption{Dependence of non-parametric tracers (from top to bottom: light concentration, asymmetry, Gini, and moment of light) on the stellar mass (left) and the environment (right) for the global sample. In each panel, we show the Spearman correlation coefficient, $\rho$,  and the corresponding probability of an uncorrelated dataset having the same distribution in parentheses (the coefficient is computed for galaxies with $M_\star>$\msun{10}).}
\label{fig:non_parametric_mass_all}
\end{figure*}

In Fig. \ref{fig:non_parametric_mass_all} we show the dependence of the non-parametric quantities studied in this manuscript as a function of stellar mass and environment for the global sample. As is the case for the parametric quantities, we find that correlations between structural parameters and stellar mass are always stronger (less likely to be random) than that found for correlations with local density.

\section{Environmental dependence for star-forming and quiescent galaxies}

\begin{figure*}
\centering
\includegraphics[width=\linewidth]{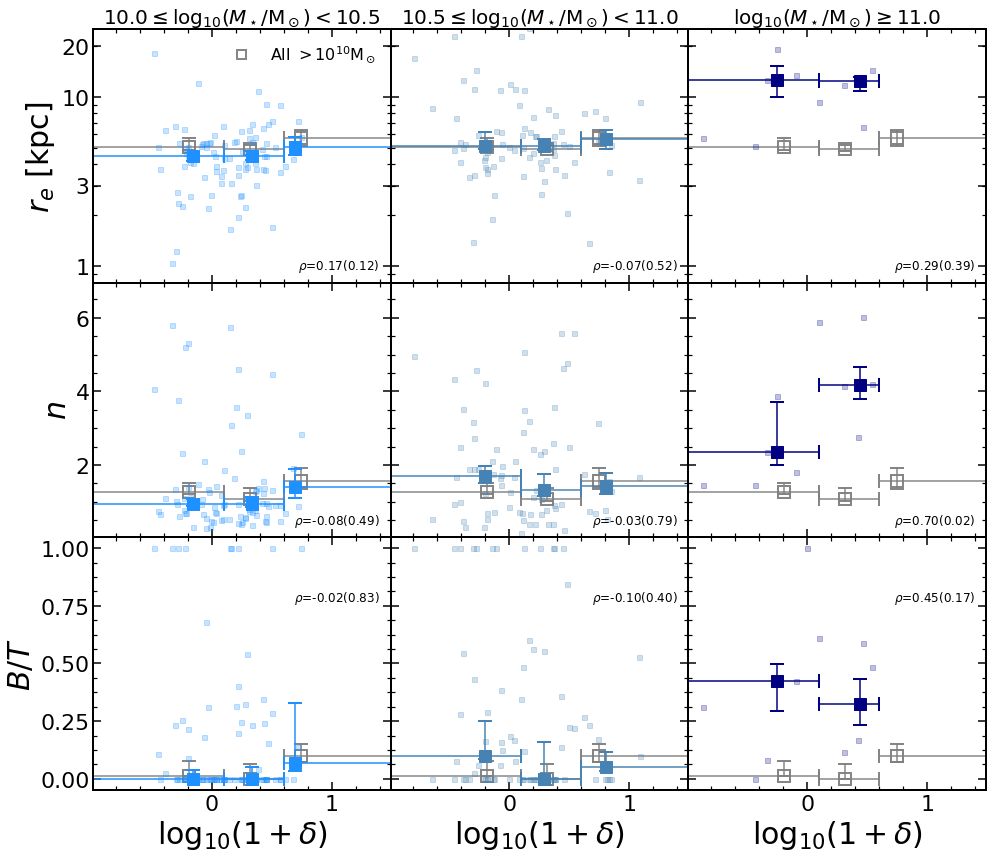}
\caption{Same as Fig. \ref{fig:parametric_mass_env}, but considering only star-forming galaxies.}
\label{fig:parametric_mass_env_SF}
\end{figure*}

\begin{figure*}
\centering
\includegraphics[width=\linewidth]{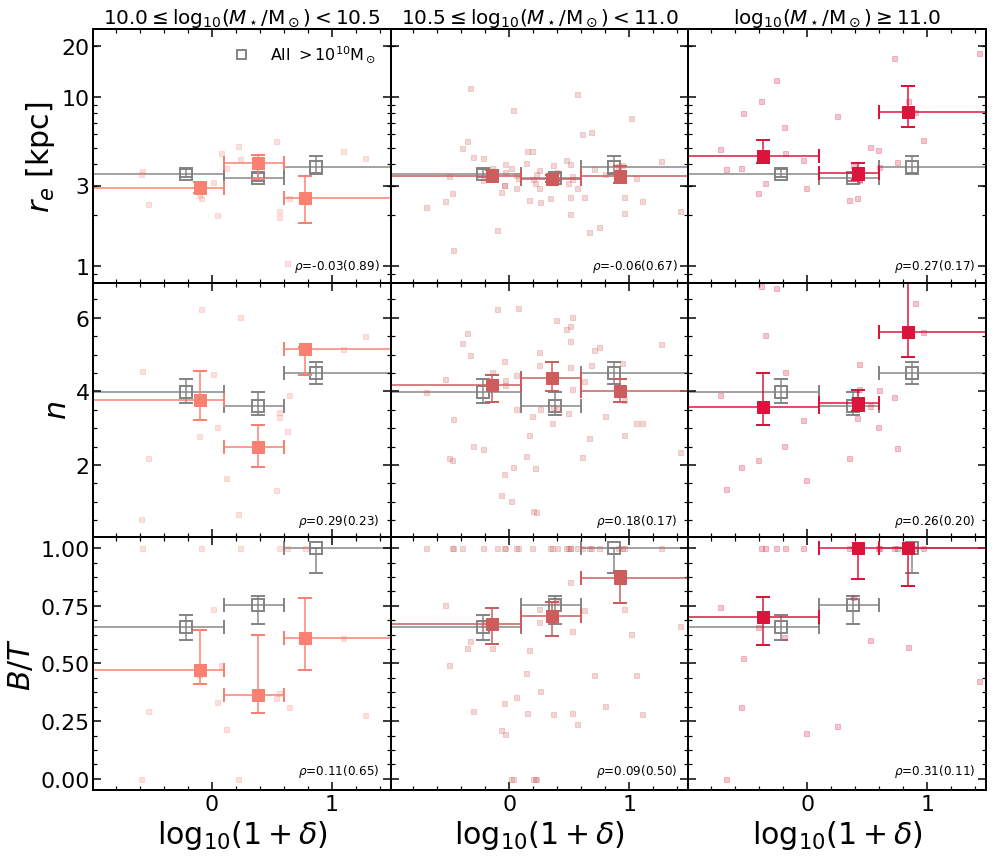}
\caption{Same as Fig. \ref{fig:parametric_mass_env}, but considering only quiescent galaxies.}
\label{fig:parametric_mass_env_Q}
\end{figure*}

\begin{figure*}
\centering
\includegraphics[width=\linewidth]{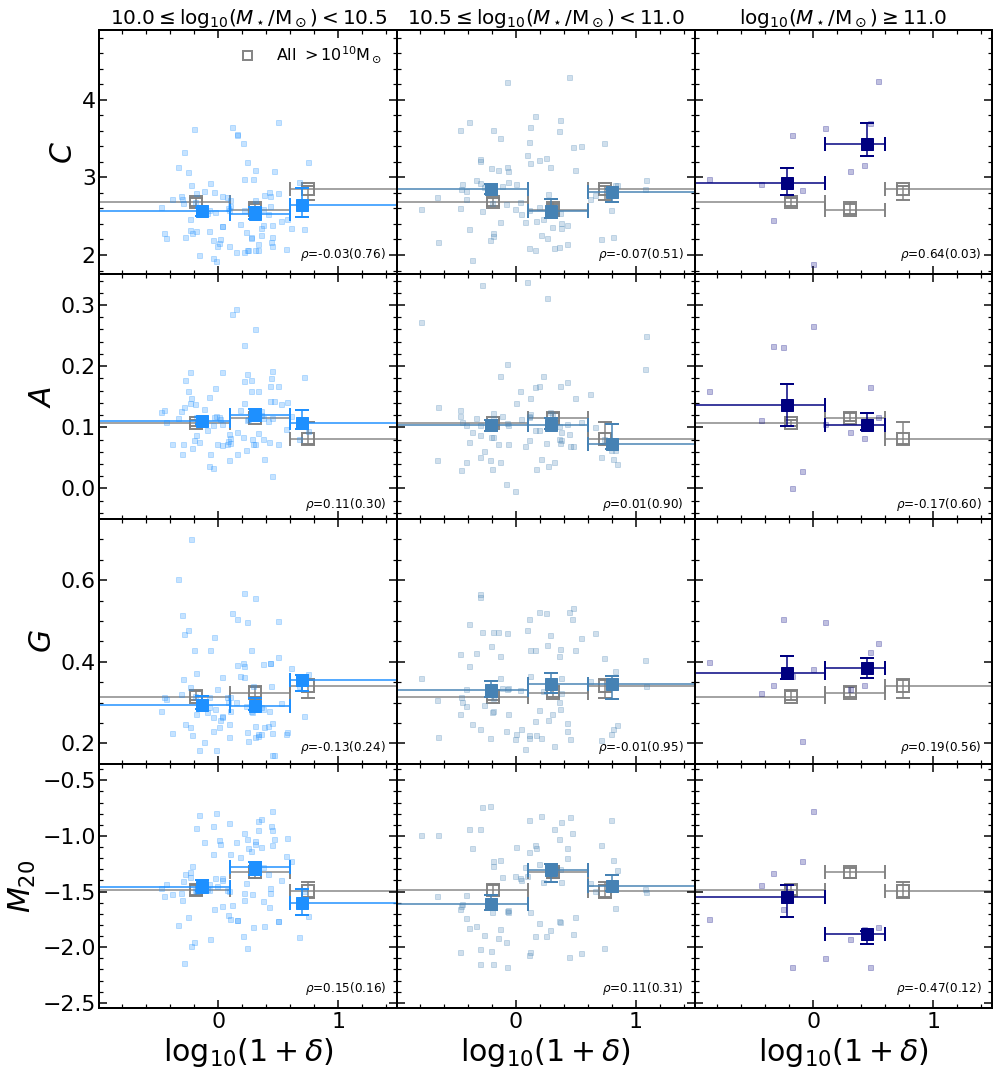}
\caption{Same as Fig. \ref{fig:nonparametric_mass_env}, but considering only star-forming galaxies.}
\label{fig:nonparametric_mass_env_SF}
\end{figure*}

\begin{figure*}
\centering
\includegraphics[width=\linewidth]{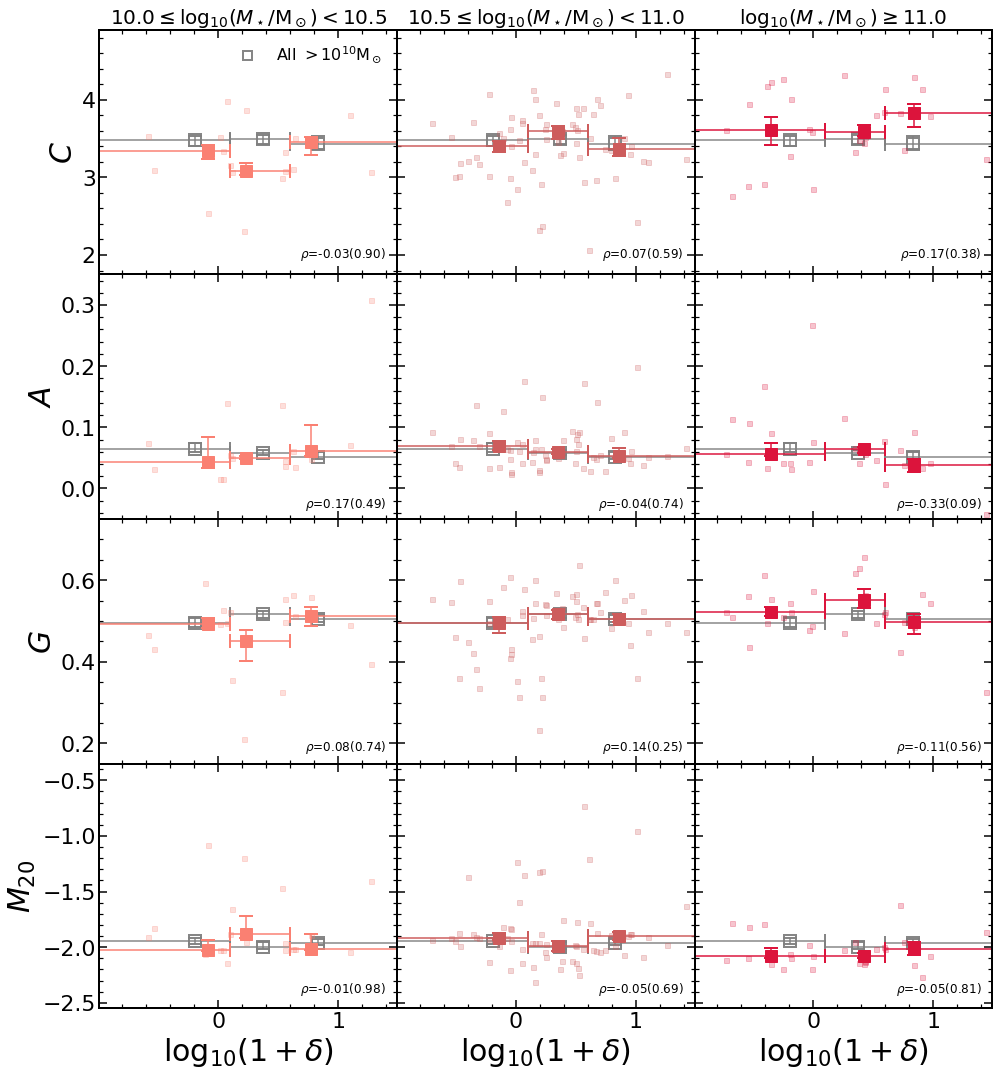}
\caption{Same as Fig. \ref{fig:nonparametric_mass_env}, but considering only quiescent galaxies.}
\label{fig:nonparametric_mass_env_Q}
\end{figure*}

In this appendix, we explore, for the sake of completeness, the relations between the structural and morphological parameters with stellar mass and local density for the populations of star-forming and quiescent galaxies. In Figs. \ref{fig:parametric_mass_env_SF} and \ref{fig:parametric_mass_env_Q} we show the relations for parametric quantities of star-forming galaxies and quiescent galaxies, respectively. We show in Figs. \ref{fig:nonparametric_mass_env_SF} and \ref{fig:nonparametric_mass_env_Q} the relation for the non-parametric quantities of star-forming galaxies and quiescent galaxies, respectively. The overall conclusion from these plots is that the correlations with local density for all presented parameters are weak at best, and non-existent in others.

\end{appendix}

\end{document}